\documentclass[prd,aps,tightenlines,superscriptaddress,showpacs,nofootinbib,eqsecnum,amsfonts,amsmath]{revtex4}

\usepackage{graphics}
\usepackage{graphicx}
\usepackage{bm}

\begin{document}

\title{
New class of post-Newtonian approximants to the waveform templates
of inspiralling compact binaries: Test-mass in the Schwarzschild spacetime}

\author{P. Ajith}\email{Ajith.Parameswaran @ aei.mpg.de}
\altaffiliation{Currently at {\it Max-Planck-Institut f\"ur Gravitationsphysik, 
Albert-Einstein-Institut, Callinstr. 38, 30167 Hannover, Germany}}
\author{Bala R. Iyer}\email{bri @ rri.res.in} 
\affiliation{Raman Research Institute, Bangalore 560 080, India}
\author{C. A. K. Robinson}\email{Craig.Robinson @ astro.cf.ac.uk}
\affiliation{School of Physics and Astronomy,Cardiff University,
5, The Parade, Cardiff,CF24 3YB, U.K.}
\author{B. S. Sathyaprakash}\email{B.Sathyaprakash @ astro.cf.ac.uk}
\affiliation{School of Physics and Astronomy,Cardiff University, 
5, The Parade, Cardiff,CF24 3YB, U.K.}

\date{\today}

\begin{abstract}
The {\it standard adiabatic} approximation to  phasing of gravitational 
waves from inspiralling compact binaries uses the post-Newtonian expansions 
of the binding energy and gravitational wave flux both truncated at the 
{\it same relative} post-Newtonian order. Motivated by the eventual need 
to go beyond the adiabatic approximation we must view the problem as the 
dynamics of the binary under conservative post-Newtonian forces and 
gravitational radiation damping. From the viewpoint of the dynamics of 
the binary, the  standard  approximation at leading order is equivalent 
to retaining the 0PN and 2.5PN terms in the acceleration and neglecting 
the intervening 1PN and 2PN terms. A complete mathematically consistent 
treatment of the acceleration at leading order should include all PN terms 
up to 2.5PN {\it without any gaps}. These define the {\it standard} and 
{\it complete} {\it non-adiabatic} approximants respectively. We  propose 
a new and simple {\it complete adiabatic} approximant constructed from the 
energy and flux functions. At the leading order it  uses the 2PN energy function
rather than the 0PN one in the standard approximation so that in spirit
it corresponds to the dynamics where there are no missing post-Newtonian 
terms in the  acceleration. We compare the overlaps of the standard and 
complete adiabatic approximant templates with the  exact waveform 
(in the adiabatic approximation) for a  test-particle orbiting a Schwarzschild 
black hole. Overlaps are computed using both the white-noise spectrum and the 
initial LIGO noise spectrum. The  complete adiabatic approximants lead to 
a remarkable improvement in the {\it effectualness} (i.e. larger overlaps 
with the exact signal) at lower PN ($<$  3PN) orders. However, standard 
adiabatic  approximants of order $\geq$ 3PN  are nearly as good as the complete 
adiabatic approximants for the construction of effectual templates. In general, 
{\it faithfulness}  (i.e. smaller biases in the estimation of parameters) of 
complete approximants is also  better than that of standard approximants.
Standard and complete approximants beyond the adiabatic approximation are 
next  studied using  the Lagrangian models of Buonanno, Chen and Vallisneri in
the test mass limit.  A limited extension  of the results  to the case of 
comparable mass binaries  is provided. In this case, standard adiabatic 
approximants achieve an effectualness  of 0.965 at order 3PN. If the comparable 
mass case is qualitatively similar to  the test mass case then  neither the 
improvement of the  accuracy of energy function from 3PN to 4PN nor the 
improvement of  the accuracy of flux function from 3.5PN to 4PN will result 
in a  significant improvement in effectualness  in the comparable mass case
for terrestrial laser interferometric gravitational wave detectors.

\end{abstract}

\maketitle
\section {Introduction}

The late-time dynamics of astronomical binaries consisting of neutron 
stars and/or black holes is dominated by relativistic motion and
non-linear general relativistic effects. The component bodies would 
be accelerated to velocities close to half the speed of light 
before they plunge towards each other, resulting in a violent event during which 
the source would be most luminous in the gravitational window. Such events are
prime targets of interferometric gravitational wave (GW) detectors like LIGO/VIRGO/GEO/TAMA that
are currently taking data at unprecedented sensitivity levels and bandwidths 
\cite{ligo,virgo,geo,tama}.

Binary coalescences are the end state of a long period of adiabatic
dynamics in which the orbital frequency of the system changes as
a result of gravitational radiation backreaction but the change in
frequency per orbit is negligible compared to the orbital frequency itself. 
Indeed, the adiabatic inspiral phase is well-modelled by the post-Newtonian (PN)
approximation to Einstein's equations but this approximation becomes
less accurate close to the merger phase. Additionally, there are different
ways of casting the gravitational wave phasing formula -- the formula that
gives the phase of the emitted gravitational wave as a function of time 
and the parameters of the system. These different approaches make use
of the post-Newtonian expansions of the binding energy and gravitational wave
luminosity of the system\footnote{In the case of binaries consisting of 
spinning bodies in eccentric orbit one additionally requires equations
describing the evolution of the individual spins and the orbital angular
momentum, but this complication is unimportant for our purposes.}.

\subsection{Standard approach to phasing formula}

The standard approach in deriving the phasing formula uses the {\it specific} 
gravitational binding energy $E(v)$ (i.e. the binding energy per unit mass) 
of the system and its luminosity ${\cal F}(v),$ both to the same relative 
accuracy \cite{CF}. Including the radiation reaction at dominant order, however, 
is not a first order correction to the dynamics of the system, rather it is a correction
that arises at ${\cal O}[(v/c)^5],$ where $v$ is the post-Newtonian expansion
parameter describing the velocity in the system and $c$ is the speed of light
\footnote{Throughout this paper we use units in which $G = c = 1$.}.
Thus, the phasing of the waves when translated to the relative motion of the
bodies implies that the dynamics is described by the dominant Newtonian force
and a correction at an order $(v/c)^5,$ but neglecting conservative force
terms that occur at orders $(v/c)^2$ and $(v/c)^4.$ Such considerations have
led to an approximation scheme in which one constructs the phasing of 
gravitational waves using the following ordinary, coupled differential equations:
\begin{equation}
\frac{d\varphi}{dt} = \frac{2v^3}{m},\ \ \ \ 
\frac{dv}{dt} = -\frac{{\cal F}(v)}{mE'(v)},
\label{eq:phasing1}
\end{equation}
where $E'(v)=dE(v)/dv$ and $m=m_1+m_2$ is the total mass of the binary.
The phasing obtained by numerically solving the above set of differential 
equations is called the {\it TaylorT1} approximant \cite{dis03}.
If the detector's motion can be neglected during the period when the
wave passes through its bandwidth then the response of the interferometer
to arbitrarily polarized waves from an inspiralling binary is given by
\begin{equation}
h(t) = \frac {4A\eta m}{D} v^2(t) \cos [\varphi(t) + \varphi_C],
\label{eq:waveform1}
\end{equation}
where $\varphi(t)$ is defined so that it is zero when the binary coalesces 
at time $t=t_C,$ $\varphi_C$ is the phase of the signal at 
$t_C,$ $\eta=m_1 m_2/m^2$ is the symmetric mass ratio, $D$ is the distance 
to the source and $A$ is a numerical constant whose value depends on the 
relative orientations of the interferometer and the binary orbit. It suffices 
to say for the present purpose that for an optimally oriented source $A=1.$

One can compute the Fourier transform $H(f)$ of the waveform given in 
Eq.~(\ref{eq:waveform1}) using the stationary phase approximation:
\begin{equation}
H(f) = \frac {4 A m^2}{D}\sqrt{\frac{5\pi \eta}{384}}  v_f^{-7/2} 
e^{i \left [2\pi f t_C - \varphi_C + \psi(f) - \pi/4 \right ]},
\label{eq:waveform2}
\end{equation}
where the phase of the Fourier transform obeys a set of differential equations given by
\begin{equation}
\frac{d\psi}{df} = 2\pi t, \ \ \ \ 
\frac{dt}{df} = - \frac{\pi m^2}{3v_f^2} \frac{E'(v_f)}{ {\cal F}(v_f)}.
\label{eq:phasing2}
\end{equation}
In the above expressions, including the post-Newtonian expansions 
of the energy and flux functions, the parameter  $v_f=(\pi m f)^{1/3}$. 
The waveform Eq.(\ref{eq:waveform2}) computed by numerically solving the 
differential equations  Eq.(\ref{eq:phasing2}) is called {\it TaylorF1} 
\cite{dis03} approximant.

Before we proceed further, let us recall the notation used in post-Newtonian
theory to identify different orders in the expansion. In the conservative
dynamics of the binary, wherein there is no dissipation, the energy is expressed
as a post-Newtonian expansion in $\epsilon = (v/c)^2,$ with the dominant
order termed Newtonian or 0PN and a correction at order $\epsilon^n = (v/c)^{2n},$ 
$n=1,2,\ldots,$ called $n$PN,  with the  dynamics involving only even powers of 
$\sqrt{\epsilon} = (v/c).$  When dissipation is added to the dynamics, then the
equation of motion will have terms of both odd and even powers of $v/c$. 
Thus, a correction of order $(v/c)^m$ is termed as $(m/2)$PN.

In the adiabatic approximation of a test-particle orbiting a Schwarzschild black hole, 
the energy function $E(v)$ is exactly calculable analytically, while the flux function 
${\cal F}(v)$ is exactly calculable numerically \cite{P93,P95,TagNak,shibata93}. 
In addition, ${\cal F}(v)$ has been calculated analytically to 5.5PN order~\cite{TTS97} 
by black hole perturbation theory~\cite{ST-LivRev}. In contrast, in the case of 
a general binary including bodies of comparable masses, the energy function 
$E(v)$ has been calculated recently to 3PN accuracy by a variety of methods 
\cite{DJS,BF,DJS02,BDE03,IF,itoh2}. The flux function ${\cal F}(v),$ on the other 
hand, has been calculated to 3.5PN accuracy 
\cite{BDIWW,BDI,WW,BIWW,B96,BIJ02,BFIJ02,ABIQ04,BDEI,BI04,BDI04} up to now only by 
the  multipolar-post-Minkowskian method and matching to a post-Newtonian 
source~\cite{Luc-LivRev}.

\subsection{Complete phasing of the adiabatic inspiral: An alternative}

The gravitational wave flux arising from the lowest order quadrupole formula,
that is the 0PN order flux, leads to an acceleration of order {\it 2.5PN} in the 
equations-of-motion. This far-zone computation of the flux requires a control of 
the dynamics, or acceleration, to only {\it Newtonian} accuracy. The lowest order 
GW phasing in the adiabatic approximation uses only the leading terms in the 
energy (Newtonian) and flux (quadrupolar) functions. For higher order phasing, 
the energy and flux functions are retained to the same relative PN orders. For 
example, at 3PN phasing, both the energy and flux functions are given to the same 
{\it relative} 3PN order beyond the leading Newtonian order. We refer to 
this usual physical treatment of the phasing of GWs computed in the adiabatic 
approximation, and used in the current LIGO/VIRGO/GEO/TAMA searches for the 
radiation from inspiralling compact binaries, as the {\it standard adiabatic} 
approximation. We will denote the $n$PN {\it standard adiabatic} approximant as 
$T(E_{[n]},{\cal F}_{n})$, where $[p]$ denotes the integer  part of $p$.

As a prelude to go beyond the standard adiabatic
approximation, let us consider the phasing of the waves in terms of
the equations of motion of the system. To this end, it is natural to order the
PN approximation in terms of its dynamics or acceleration. From 
the viewpoint of the dynamics, the leading order standard adiabatic 
approximation is equivalent to using the 0PN (corresponding to
0PN conserved energy) and 2.5PN (corresponding to the Newtonian or 0PN
flux) terms in the acceleration ignoring the intervening 1PN and 2PN terms.
A complete, mathematically  consistent treatment of the acceleration, 
however, should include {\it all} PN  terms in the acceleration up to 2.5PN,
{\it without any gaps}. We refer to the dynamics of the binary, 
and the resulting waveform, arising from the latter as the 0PN {\it complete 
non-adiabatic} approximation. In  contrast, the waveform arising from
the former choice, with  gaps in the acceleration at 1PN and 2PN,
is referred to as the 0PN  {\it standard non-adiabatic} approximation.
Extension to higher-order phasing is obvious. At 1PN the standard non-adiabatic 
approximation would involve acceleration terms at orders 0PN, 1PN, 2.5PN and 3.5PN,
whereas the complete non-adiabatic approximation would additionally involve
the 2PN and 3PN acceleration terms.

Finally, we propose a simple extension of the above construction
to generate a new class of approximants in the adiabatic regime.
To understand the construction let us examine the lowest
order case. Given the 0PN flux (leading to an acceleration at 2.5PN),
one can choose the energy function at 2PN (equivalent to 2PN conservative 
dynamics) instead of the standard choice 0PN (equivalent to 0PN or Newtonian 
conservative dynamics). This is the adiabatic analogue
of the complete non-adiabatic approximant\footnote{
In this case one may also choose the energy function to 3PN accuracy and 
construct a complete approximant leading to 3PN acceleration}.
Extension to higher PN orders follows naturally. For instance,
corresponding to the flux function at 1PN (1.5PN),
the dissipative force is at order 3.5PN (4PN), and, therefore, 
the conservative dynamics, and the associated energy
function, should be specified up to order 3PN (4PN).
In general, given the flux at $n$PN-order, a corresponding complete
adiabatic approximant is constructed by choosing the 
energy function at order $[n+2.5]$PN, where as mentioned before,
$[p]$ denotes the integer  part of $p$. We refer to the dynamics of the binary 
and the resulting waveform arising from such considerations, as the 
{\it complete adiabatic} approximation. We will denote the $n$PN 
{\it complete adiabatic} approximant as $T(E_{[n+2.5]},{\cal F}_{n})$.

Before moving ahead the following point is worth emphasizing:
The standard adiabatic phasing is, by construction, consistent
in the {\it relative PN} order of its constituent energy and flux
functions, and thus {\it unique} in its ordering of the PN terms. 
Consequently, one can construct different inequivalent,
but consistent, approximations as discussed in Ref.~\cite{dis03} by
choosing to retain the involved functions or re-expand them.
The complete adiabatic phasing, on the other hand, is constructed
so that it is consistent in spirit  with the underlying dynamics,
or acceleration, rather than with the relative PN orders of the
energy and flux functions. Consequently, it has a unique meaning
only when the associated energy and flux functions are used 
{\it without any further re-expansions} when working out the
phasing formula. As a result, though the complete non-adiabatic
approximant is more consistent than the standard non-adiabatic
approximant in treating the PN accelerations, in the adiabatic case 
there is no rigorous sense in which one can claim that either of the 
approximants is more consistent than the other. The important point,
as we shall see is that, not only are the two approximants {\it not} 
the same but the new complete adiabatic approximants are
closer to the exact solution than the standard adiabatic approximants.

In our view, these new approximants  should be of some interest.
They are simple generalizations of the {\it standard adiabatic}
approximants coding information of the PN dynamics beyond the standard
approximation without the need for numerical integration 
of the equations of motion. They should be appropriate approximants to focus
on when one goes beyond the adiabatic picture and investigates the
differences  stemming from the use of more  complete
equations of motion (see Section \ref{sec:non-adiabatic models}).

In the case of comparable mass binaries, the energy function is currently known
up to 3PN order and hence it would be possible to compute the 
complete adiabatic phasing of the waves to only 1PN order. One is thus obliged in 
practice to follow the standard adiabatic approximation to obtain the 
phasing up to 3.5PN order.  Consequently, it is a 
relevant question to ask how `close' are the complete and standard adiabatic 
approximants. The standard adiabatic approximation would be justified 
if we can verify that it produces in most cases a good lower bound to 
the mathematically consistent, but calculationally more demanding, 
complete adiabatic approximation. In this paper we compare 
the standard and complete models by explicitly 
studying their overlaps with the exact waveform which can be computed in
the adiabatic approximation of a test mass motion in a Schwarzschild 
spacetime. The availability, in this case, of the exact (numerical) 
and approximate (analytical) waveforms to as high a PN order as 
$(v/c)^{11}$, allows one to investigate the issue exhaustively, and 
provides the main motivation for the present analysis. Assuming that 
the comparable mass case is qualitatively similar and a simple 
$\eta$-distortion of the test mass case would then provide a 
plausible justification for the standard adiabatic treatment of the GW 
phasing employed in the literature\footnote{Note, however, that the view that the
comparable mass case is just a $\eta$-distortion of the test mass
approximation is not universal. In particular, Blanchet \cite{blanchet01}
has argued that the dynamics of a binary consisting of two bodies of comparable masses is
very different from, and possibly more accurately described by post-Newtonian
expansion than, the test mass case.}.
\begin{figure}[t]
\centering \includegraphics[width=5in]{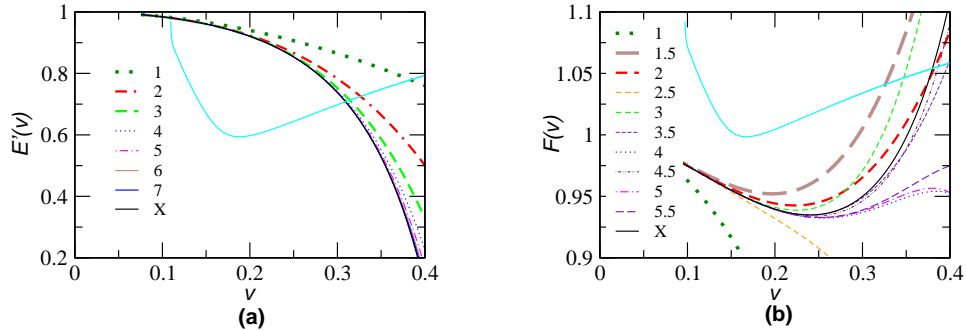}
\caption{Various T-approximants of Newton-normalized ($v$-derivative 
of) energy function $E'_{\rm T}(v)/E'_{\rm N}(v)$ (left), and flux 
function ${\cal F}_{\rm T}(v)/{\cal F}_{\rm N}(v)$ (right) in the 
test mass limit along with the exact functions (denoted by $X$). Also 
plotted is the amplitude spectral density (per $\sqrt{\rm Hz}$) of 
initial LIGO noise in arbitrary units.}
\label{fig:TMEnFlfnsLigo}
\end{figure}

\subsection{Non-adiabatic inspiral}
The phasing formulas derived under the various adiabatic approximation schemes
assume that the orbital frequency changes slowly over each orbital period.
In other words, the change in frequency $\Delta \nu $ over one orbital period $P$
is assumed to be much smaller than the orbital frequency $\nu \equiv P^{-1}.$
Denoting by $\dot \nu $ the time-derivative of the frequency, the adiabatic 
approximation is equivalent to the assumption that $\Delta \nu  = \dot \nu  P \ll \nu $
or $\dot \nu /\nu ^2 \ll 1.$
This assumption becomes somewhat weaker, and it is unjustified to use
the approximation $\dot E = -{\cal F},$ when the two bodies are quite
close to each other. Buonanno and Damour \cite{bd99,bd00} introduced a non-adiabatic
approach to the two-body problem called the {\it effective one-body} (EOB)
approximation.  In this approximation one solves for the relative 
motion of the two bodies using an effective Hamiltonian with a dissipative
force put-in by hand.  EOB allows to extend the dynamics beyond the adiabatic
regime, and the last stable orbit, into the plunge phase of the 
coalescence of the two bodies \cite{bd00,DGG,GGB1,GGB2}.  

Recently, Buonanno, Chen and Vallisneri \cite{BCV02}
have studied a variant of the non-adiabatic model but using the
effective Lagrangian constructed in the post-Newtonian approximation.
We shall use both the {\it standard} and {\it complete} non-adiabatic
Lagrangian models in this study and see how they converge to the 
exact waveform defined using the adiabatic approximation\footnote{See
Sec. \ref{sec:non-adiabatic models} for a caveat in this approach.}.

\subsection{What this study is about}

In our study we will use the {\it effectualness} and {\it faithfulness} 
(see below) to quantify  how good the various approximation schemes are.
There are at least three different contexts in which one can examine the 
performance of an approximate template family  relative to an exact one. 
Firstly, one can think of a mathematical family of approximants and examine its 
convergence towards some exact limit. Secondly, one can ask whether 
this mathematical family of approximants correctly represents the GWs from 
some physical system. Thirdly, how does this family of approximate templates 
converge to the exact solution in the sensitive bandwidth of a particular 
GW detector. In the context of GW data analysis, the third context will be 
relevant and studied in this paper. Although there is no direct application 
to GW data analysis, equally interesting is the mathematical question 
concerning the behavior of different approximations, and the waveforms they 
predict, in the strongly non-linear regime of the dynamics of the binary, 
which is also studied in this paper. The latter obviously does not require 
the details of the detector-sensitivity and it is enough to study the problem 
assuming a flat power spectral density (i.e. a {\it white-noise} background) 
for the detector noise.   

To summarize, our approach towards the problem will be two-pronged. 
First, we will study the problem as a general mathematical question 
concerning the nature of templates defined using PN approximation 
methods. We shall deal with four families of PN templates -- the 
standard adiabatic, complete adiabatic, standard non-adiabatic and 
complete non-adiabatic (in particular, Lagrangian-based) approximants --
and examine their closeness, defined by using effectualness and
faithfulness, to the exact waveform defined in the adiabatic approximation. 
Since this issue is a general question independent of the characteristics of 
a particular GW detector, we first study the problem assuming the white-noise 
case. Having these results, we then proceed to see how and which of these 
results are applicable to a specific detector, namely the initial LIGO  
\footnote{The one-sided noise power spectral density (PSD) of the initial LIGO is given in
terms of a dimensionless frequency $x=f/f_0$ by~\cite{dis03}, 
$S_h(x) = 9 \times 10^{-46} \left [(4.49x)^{-56} + 0.16 x^{-4.52} + 0.52 + 0.32 x^2\right ]$
where $f_0=150$ Hz and the PSD rises steeply below a lower cut-off $f_c = 40$ Hz.}.
During the course of this study, we also attempt to assess the 
relative importance of  improving the accuracy of the energy 
and flux functions by studying the overlaps of the PN templates constructed from 
different orders of energy and flux functions with the exact waveform. It should be  
kept in mind that this work is  solely restricted to the inspiral part of the signal 
and neglects the plunge and quasi-normal mode ringing phases of the 
binary~\cite{FH98,bd00,dis03,BCV02,DIJS,QNM}.

\subsection {Effectualness and Faithfulness}
In order to measure the accuracy of our approximate
templates we shall compute their overlap with a fiducial exact signal.
We shall consider two types of overlaps~\cite{dis01,dis02,dis03,dis04}.
The first one is the
{\it faithfulness} which is the overlap of the approximate template 
with the exact signal computed by keeping
the intrinsic parameters (e.g. the masses of the two bodies)
of both the template and the signal to be the same but maximizing
over the extrinsic (e.g. the time-of-arrival and the phase at that
time) parameters. The second one is the {\it effectualness} which is
the overlap of the approximate template with the exact signal computed
by maximizing the overlap over both the intrinsic and extrinsic parameters.
Faithfulness is a measure of how good is the template waveform in both
detecting a signal and measuring its parameters. However, effectualness
is aimed at finding whether or not an approximate template model is
good enough in detecting a signal without reference to its use in 
estimating the parameters.
As in previous studies, we take overlaps greater than 96.5\% 
to be indicative of a good approximation.  

\subsection {Organization of the paper}

In the next section we study the test-mass waveforms in the adiabatic 
approximation. We discuss the construction of the exact energy and flux 
functions as well as the respective T-approximants. The overlaps of various 
standard adiabatic and complete adiabatic approximants are also compared 
in this Section. Section \ref{sec:non-adiabatic models} deals with the 
non-adiabatic approximation. Section \ref{sec:compmass} explores the 
extension of the results in the comparable mass case. It presents the 
energy and flux functions which are the crucial inputs for the construction 
of the fiducial `exact' waveform as well as the approximate waveforms 
followed by a discussion of the results. In the last section we summarize 
our main conclusions.

One of the  main conclusions  of this paper  is that the effectualness of the test-mass 
approximants significantly improves in the complete adiabatic  approximation at PN 
orders below 3PN. However, standard adiabatic  approximants of order $\geq$ 3PN  are 
nearly as good as  the complete adiabatic approximants for the construction of 
effectual templates. In the comparable mass case the problem can be only
studied at the lowest two PN orders. No strong conclusions can be drawn as in the test 
mass case. Still, the trends indicate that the standard adiabatic approximation
provides a good lower bound to the complete adiabatic approximation for the construction 
of both effectual and faithful templates at PN orders $\geq$ 1.5PN. From the detailed 
study of test-mass templates we also conclude that, provided the comparable mass case 
is qualitatively similar to the test mass case, neither the improvement of the accuracy 
of energy function from 3PN to 4PN nor the improvement of the accuracy of flux function 
from 3.5PN to 4PN will result in a significant improvement in effectualness in the comparable 
mass case. As far as faithfulness is concerned, it is hard to reach any conclusion. 
To achieve the target sensitivity of 0.965 in effectualness corresponding to
a 10\% loss in the event-rate, standard adiabatic approximants of order 2PN and 3PN 
are required for the $(10 M_\odot,10 M_\odot)$ and $(1.4 M_\odot,1.4 M_\odot)$ 
binaries, respectively, when restricting to only the inspiral phase. (Be warned that 
this is not a good approximation in the BH-BH case since the approach to the plunge 
and merger lead to significant differences.)

\section {Test mass waveforms in the adiabatic approximation}
\label{sec:testmass}

\begin{figure}[t]
\centering \includegraphics[width=5in]{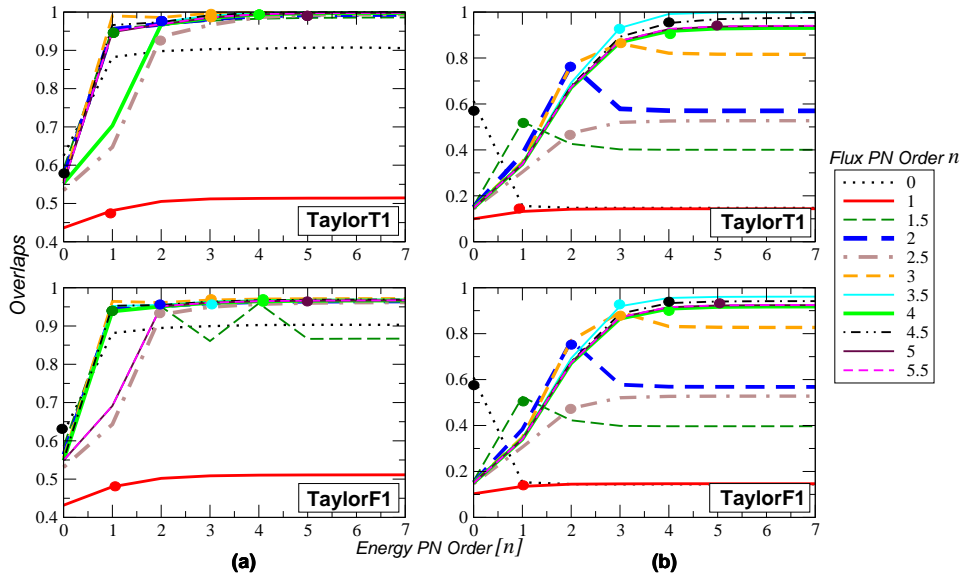}
\caption{Effectualness (left panels) and faithfulness (right panels) 
of various test mass {\it TaylorT1} and {\it TaylorF1} templates in 
detecting a signal from a $(1 M_\odot,10 M_\odot)$ binary in 
white-noise.  Different lines in the panels correspond to different 
orders of the flux function. Each line shows how the overlaps are 
evolving as a function of the accuracy of the energy function. 
Standard adiabatic  approximants $T(E_{[n]},{\cal F}_{n})$ are marked with thick 
dots.}
\label{fig:systemB-WN}
\end{figure}
 
Our objective is to compare the {\it effectualness} (i.e larger overlaps 
with the exact signal) and {\it faithfulness} (i.e. smaller bias in the 
estimation of parameters) of the standard adiabatic $T(E_{[n]},{\cal F}_{n})$ 
and complete adiabatic $T(E_{[n+2.5]},{\cal F}_{n})$ approximants. As a 
by-product of this study, we would also like to have an understanding 
of the the relative importance of improving the accuracy of the 
energy function and flux function. Thus, what we will do is to take 
all possible combinations of T-approximants\footnote{We follow \cite{dis03}
in denoting the  precise scheme used for constructing the approximant.}
of energy and flux functions, construct PN templates and calculate the overlap 
of these templates with the exact waveform. In all cases, the exact waveform is 
constructed by numerically integrating the phasing formula in the time-domain 
[{\it TaylorT1} approximant, cf. Eqs.~(\ref{eq:phasing1}) and (\ref{eq:waveform1})]. 
The waveforms (both the exact and approximate) are all terminated at $v_{\rm lso}
=1/\sqrt{6},$ which corresponds to $F_{\rm lso}\simeq43$ Hz for the 
$(1M_{\odot},100M_{\odot})$ binary, $F_{\rm lso}\simeq86$ Hz for the 
$(1M_{\odot},50M_{\odot})$ binary and $F_{\rm lso}\simeq399$ Hz for the 
$(1M_{\odot},10M_{\odot})$ binary~\footnote{Here, $v_{\rm lso}$ is the velocity 
at the last stable circular orbit of Schwarzschild geometry having the same mass 
as the total mass $m_1+m_2$ of the binary (we adopt units in which $c=G=1$) and 
$F_{\rm lso}$ is the GW frequency corresponding to it.}. The lower frequency 
cut-off of the waveforms is chosen to be $F_{\rm low}= 20$ Hz. 

In this study, we restrict to approximants TaylorT1 and TaylorF1 
since they do not involve any further re-expansion in
the phasing formula and hence there is no ambiguity when we construct
the phasing of the waves using approximants with unequal orders 
of the energy and flux functions.

\subsection {The energy function}
\begin{figure}[t]
\centering \includegraphics[width=5in]{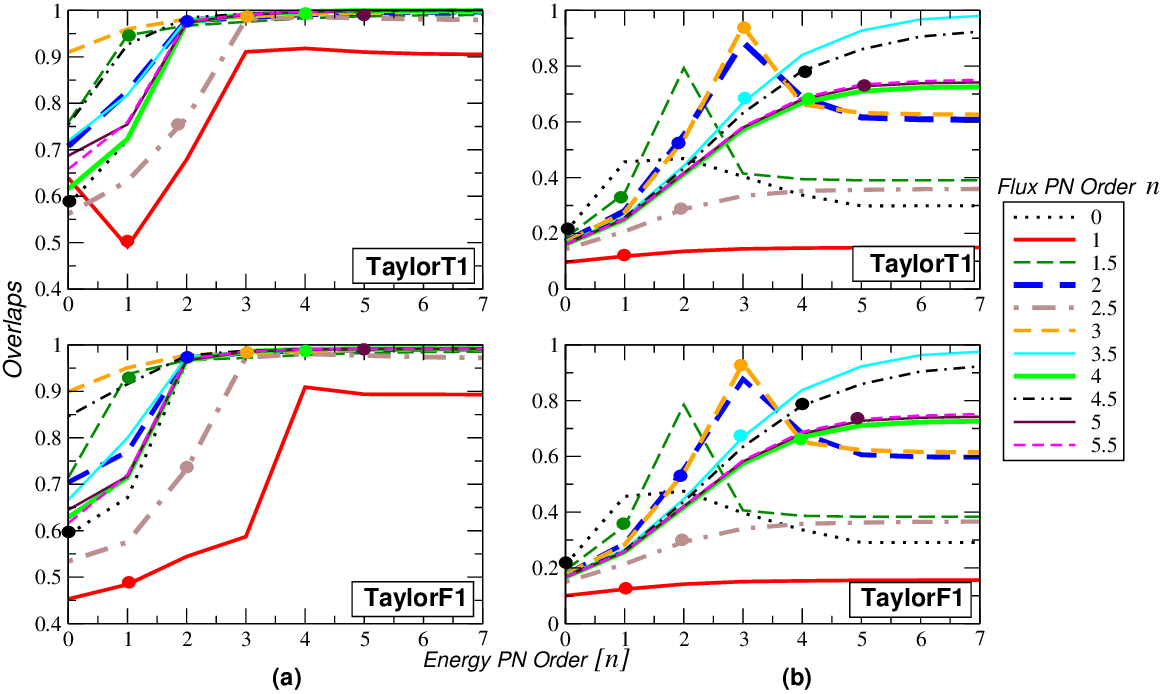}
\caption{As in Fig.~\ref{fig:systemB-WN} except that the signal 
corresponds to a $(1 M_\odot,50 M_\odot)$ binary.}
\label{fig:systemD-WN}
\end{figure}
In the case of a test-particle $m_2$ moving in circular orbit in the 
background of a Schwarzschild black hole of mass $m_1$, where 
$m_2/m_1 \rightarrow 0$, the energy function $E(x)$
in terms of the invariant argument $x\equiv v^2$
is given by
\begin{equation}
E_{\rm exact}(x) =\eta \,
\frac{1-2x}{\sqrt{1-3x}}\,,
\label{eq:Eexact}
\end{equation}
The associated $v$-derivative entering the phasing formula is 
\begin{equation}
E_{\rm exact}'(v)= 2 v\left.\frac {d E(x)}{dx} \right|_{x=v^2}=
-\eta v\,\frac {(1-6\,v^2)}{(1-3\,v^2)^{3/2}}\,. 
\label{eq:Eprimeexact}
\end{equation}
We use the above exact $E'(v)$ to construct the exact waveform in the 
test-mass case. In order to construct various approximate PN 
templates, we Taylor-expand $E_{\rm exact}'(v)$ and truncate it at 
the necessary orders.
\begin {eqnarray}
E'_{7PN}(v) & = & -\eta v\left[1-\frac{3 \, v^2 }{2}-\frac{81 \, 
v^4}{8}-\frac{675 \, v^6}{16}-\frac{19845 \, v^8}{128}\right. 
\nonumber\\
& - &\left.\frac{137781 \, v^{10}}{256}- \frac{1852389 \, v^{12}}{1024}- 
\frac {12196899 \, v^{14}}{2048} + {\cal O}(v^{16})\right] .
\label {eq:E_taylor}
\end{eqnarray}
Different T-approximants of the energy function $E'_T(v)$ along with 
$E'_{\rm exact}(v)$ are plotted in Fig.~\ref{fig:TMEnFlfnsLigo}$a$.

\subsection {The flux function}
In the test-particle limit, the exact gravitational-wave flux 
has been computed numerically with good accuracy \cite{P95}. We will use this 
flux function (see Fig.~\ref{fig:TMEnFlfnsLigo}$b$), along with the 
energy function given by Eq.~(\ref{eq:Eprimeexact}), to construct an 
exact waveform in the test-mass case. 
In the test-particle limit, the GW  flux is also known analytically
to 5.5PN order from black hole perturbation theory \cite{TTS97} and  given by
\begin{equation}
{\cal F}(v ) = \frac{32}{5} \eta^2 v^{10} 
\left[\sum_{k=0}^{11}A_k v^k +
\left ( \sum_{m=6}^{11} B_m v^m  \right ) \ln v + {\cal O} \,(v^{12}) \, \right],
\label{eq:exactflux}
\end{equation}
where the various coefficients $A_k$ and $B_k$ are \cite{TTS97},
\allowdisplaybreaks{\begin{eqnarray}
A_0    & = &1\, ,\ \ A_1  = 0\, , \ \ A_2 = -\frac{1247}{336}\, , \ \ A_3 = 4 \,\pi\, ,\ \ 
A_4    = -\frac{44711}{9072}\, , \ \ A_5  = -\frac{8191 \, \pi}{672}\, , \nonumber\\
A_6    & = &\frac{6643739519}{69854400} + \frac{16 \, \pi^2}{3} - \frac{1712 \,\gamma}{105} 
         - \frac{1712\,{\rm ln}\, 4}{105} \, ,  \ \ 
A_7    = -\frac{16285\, \pi}{504}\, ,  \nonumber\\
A_8    & = &-\frac{323105549467}{3178375200} + \frac{232597 \, \gamma}{4410} - \frac{1369 \, \pi^2}{126} 
+ \frac{39931 \,{\rm ln}\, 2}{294} - \frac{47385\,{\rm ln}\, 3}{1568}\,, \nonumber\\
A_9    & = & \frac{265978667519 \,  \pi}{745113600} - \frac{6848 \,\gamma \, \pi}{105} 
	 - \frac{13696 \, \pi \, {\rm ln}\, 2 }{105}\,, \nonumber\\
A_{10} & = &-\frac{2500861660823683}{2831932303200} + \frac{916628467 
\, \gamma}{7858620} 
	 - \frac{424223 \, \pi^2}{6804}	- \frac{83217611 \,{\rm ln}\, 2}{1122660} + \frac{47385 \, 
{\rm ln}\, 3}{196}\,,\nonumber\\
A_{11} & = & \frac{8399309750401 \, \pi}{101708006400} + \frac{177293 \, \gamma \,\pi}{1176} 
	 + \frac{8521283 \, \pi \,{\rm ln}\, 2}{17640} - \frac{142155 \,\pi \,{\rm ln}\, 3}{784}\,,  \ \ 
B_6    = - \frac{1712}{105}\, ,\ \ 
\nonumber\\
B_7 & = & 0\, ,\ \ 
B_8    = \frac{232597}{4410} \, ,\ \ 
B_9    = \frac{-6848 \, \pi }{105}\, ,\ \ 
B_{10} = \frac{916628467}{7858620},\ \ 
B_{11} = \frac{177293 \, \pi}{1176}.  
\label{eq:exactfluxcoeffs}
\end{eqnarray}
}
We will use the energy and flux functions given by 
Eq.~(\ref{eq:E_taylor}) -- Eq.~(\ref{eq:exactfluxcoeffs}) to 
construct various approximate templates by truncating the expansions 
at the necessary order. The different T-approximants of the flux 
function ${\cal F}_{\rm T}(v )$ along with the (numerical) exact flux 
 ${\cal F}_{\rm exact}(v )$ are plotted in 
Fig.~\ref{fig:TMEnFlfnsLigo}$b$.
\begin{figure}[t]
\vskip 0.5 true cm
\centering \includegraphics[width=5in]{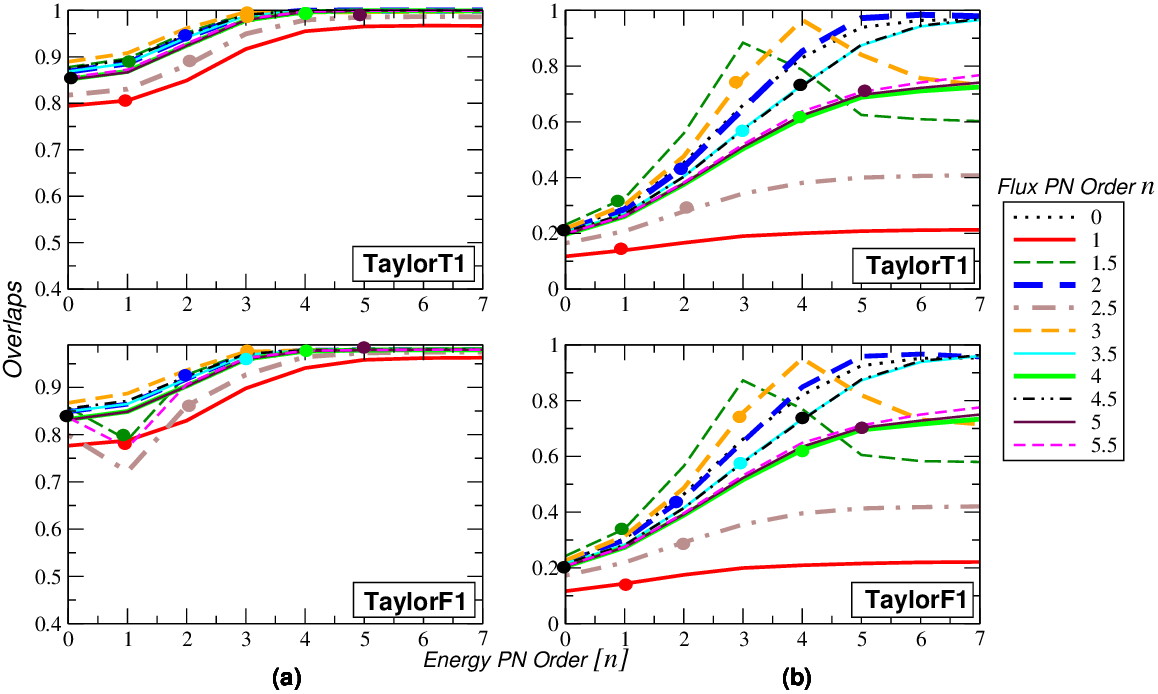}
\caption{As in Fig.~\ref{fig:systemB-WN} except that the signal 
corresponds to a $(1 M_\odot,100 M_\odot)$ binary.}
\label{fig:systemE-WN}
\end{figure}

\subsection{ Comparison of standard and complete adiabatic approximants}

We present the results of our study in the test mass limit in four parts. 
In the first part we discuss our conclusions on the mathematical problem 
of the closeness of the standard adiabatic and complete adiabatic template 
families with the  family of exact waveforms in the adiabatic approximation. 
In the next part  we exhibit  our results in   the case of the initial 
LIGO detector. In the third part we compare the relative importance of 
improving the accuracy of the energy and flux functions. Finally, in the 
fourth part we compare the total number of GW cycles and the number of useful 
cycles accumulated by various standard adiabatic  and complete adiabatic 
approximants. 

\subsubsection{ White-noise case}
\label{sec:adibatic-SC-TM-wn}
\begin{table} 
\caption{Effectualness of {\it standard adiabatic} $T(E_{[n]},{\cal F}_{n})$ and 
{\it complete adiabatic} $T(E_{[n+2.5]},{\cal F}_{n})$ templates in the test mass 
limit. Overlaps are calculated assuming a flat spectrum for the detector noise 
(white-noise).}
\begin{tabular}{ccccccccccccccccccc}
\hline
\hline
&\vline&\multicolumn{5}{c}{$(1M_{\odot},10M_{\odot})$} &\vline & \multicolumn{5}{c}
{$(1M_{\odot},50M_{\odot})$} &\vline & \multicolumn{5}{c}{$(1M_{\odot},100M_{\odot})$} \\
\cline{2-19}
&\vline&\multicolumn{2}{c}{{\it TaylorT1}}&\vline&\multicolumn{2}{c}{{\it TaylorF1}}
&\vline&\multicolumn{2}{c}{{\it TaylorT1}}&\vline&\multicolumn{2}{c}{{\it TaylorF1}}
&\vline&\multicolumn{2}{c}{{\it TaylorT1}}&\vline&\multicolumn{2}{c}{{\it TaylorF1}}\\
\cline{2-19}
Order ($n$)&\vline& {\it S} & {\it C} &\vline&  {\it S} & {\it C} 
&\vline& {\it S} & {\it C} &\vline&  {\it S} & {\it C} &\vline& 
{\it S} & {\it C} &\vline&  {\it S} & {\it C}\\
\hline
\hline
0PN&\vline& 0.6250 & 0.8980 &\vline& 0.6212 & 0.8949 &\vline& 0.5809 & 0.9726 &\vline& 
0.5917 & 0.9644 &\vline& 0.8515 & 0.9231 &\vline& 0.8318 & 0.9017\\
1PN&\vline& 0.4816 & 0.5119 &\vline& 0.4801 & 0.5086 &\vline& 0.4913 & 0.9107 &\vline& 
0.4841 & 0.5871 &\vline& 0.8059 & 0.9169 &\vline& 0.7874 & 0.8980\\
1.5PN&\vline& 0.9562 & 0.9826 &\vline& 0.9448 & 0.9592 &\vline& 0.9466 & 0.9832 &\vline& 
0.9370 & 0.9785 &\vline& 0.8963 & 0.9981 &\vline& 0.7888 & 0.9788\\
2PN&\vline& 0.9685 & 0.9901 &\vline& 0.9514 & 0.9624 &\vline& 0.9784 & 0.9917 &\vline& 
0.9719 & 0.9872 &\vline& 0.9420 & 0.9993 &\vline& 0.9178 & 0.9785\\
2.5PN&\vline& 0.9362 & 0.9924 &\vline& 0.9298 & 0.9602 &\vline& 0.7684 & 0.9833 &\vline& 
0.7326 & 0.9772 &\vline& 0.8819 & 0.9858 &\vline& 0.8610 & 0.9730\\
3PN&\vline& 0.9971 & 0.9991 &\vline& 0.9677 & 0.9713 &\vline& 0.9861 & 0.9946 &\vline& 
0.9821 & 0.9886 &\vline& 0.9965 & 0.9959 &\vline& 0.9756 & 0.9792\\
3.5PN&\vline& 0.9913 & 0.9996 &\vline& 0.9636 & 0.9688 &\vline& 0.9902 & 0.9994 &\vline& 
0.9858 & 0.9914 &\vline& 0.9885 & 1.0000 &\vline& 0.9690 & 0.9800\\
4PN&\vline& 0.9937 & 0.9973 &\vline& 0.9643 & 0.9663 &\vline& 0.9975 & 0.9996 &\vline& 
0.9903 & 0.9914 &\vline& 0.9968 & 0.9992 &\vline& 0.9769 & 0.9795\\
4.5PN&\vline& 0.9980 & 0.9999 &\vline& 0.9671 & 0.9690 &\vline& 0.9967 & 1.0000 &\vline& 
0.9902 & 0.9913 &\vline& 0.9996 & 1.0000 &\vline& 0.9787 & 0.9801\\
5PN&\vline& 0.9968 & 0.9979 &\vline& 0.9661 & 0.9667 &\vline& 0.9994 & 0.9994 &\vline& 
0.9913 & 0.9914 &\vline& 0.9992 & 0.9991 &\vline& 0.9790 & 0.9797\\
\hline
\hline
\label{table:Effectualness-SC-WN}
\end{tabular}
\end {table}	

\begin{table} 
\caption{Faithfulness of {\it standard adiabatic} $T(E_{[n]},{\cal F}_{n})$ and {\it complete adiabatic}  
$T(E_{[n+2.5]},{\cal F}_{n})$ templates in the test mass limit. Overlaps are calculated 
assuming a flat spectrum for the detector noise (white-noise).}
\begin{tabular}{ccccccccccccccccccc}
\hline
\hline
&\vline&\multicolumn{5}{c}{$(1M_{\odot},10M_{\odot})$} &\vline & \multicolumn{5}{c}
{$(1M_{\odot},50M_{\odot})$} &\vline & \multicolumn{5}{c}{$(1M_{\odot},100M_{\odot})$} \\
\cline{2-19}
&\vline&\multicolumn{2}{c}{{\it TaylorT1}}&\vline&\multicolumn{2}{c}{{\it TaylorF1}}
&\vline&\multicolumn{2}{c}{{\it TaylorT1}}&\vline&\multicolumn{2}{c}{{\it TaylorF1}}
&\vline&\multicolumn{2}{c}{{\it TaylorT1}}&\vline&\multicolumn{2}{c}{{\it TaylorF1}}\\
\cline{2-19}
Order ($n$) &\vline& {\it S} & {\it C} &\vline&  {\it S} & {\it C} 
&\vline& {\it S} & {\it C} &\vline&  {\it S} & {\it C} &\vline& {\it S}
 & {\it C} &\vline&  {\it S} & {\it C}\\
\hline
\hline

0PN &\vline& 0.6124 & 0.1475 &\vline& 0.6088 & 0.1446 &\vline& 0.2045 & 0.4683 &\vline& 
0.2104 & 0.4750 &\vline& 0.2098 & 0.4534 &\vline& 0.2208 & 0.4641 \\
1PN &\vline& 0.1322 & 0.1433 &\vline& 0.1350 & 0.1461 &\vline& 0.1182 & 0.1446 &\vline& 
0.1236 & 0.1508 &\vline& 0.1395 & 0.1901 &\vline& 0.1432 & 0.1994  \\
1.5PN &\vline& 0.5227 & 0.4005 &\vline& 0.5241 & 0.3967 &\vline& 0.3444 & 0.3947 &\vline&
 0.3505 & 0.3866 &\vline& 0.3260 & 0.7869 &\vline& 0.3399 & 0.7700 \\
2PN &\vline& 0.7687 & 0.5707 &\vline& 0.7680 & 0.5689 &\vline& 0.5518 & 0.6871 &\vline& 
0.5535 & 0.6827 &\vline& 0.4377 & 0.8528 &\vline& 0.4506 & 0.8486 \\
2.5PN &\vline& 0.4735 & 0.5268 &\vline& 0.4748 & 0.5278 &\vline& 0.2874 & 0.3561 &\vline& 
0.2933 & 0.3625 &\vline& 0.2787 & 0.4001 &\vline& 0.2918 & 0.4133 \\
3PN &\vline& 0.8629 & 0.8165 &\vline& 0.8932 & 0.8277 &\vline& 0.9420 & 0.6317 &\vline& 
0.9334 & 0.6222 &\vline& 0.7579 & 0.8407 &\vline& 0.7570 & 0.8194 \\
3.5PN &\vline& 0.9309 & 0.9979 &\vline& 0.9194 & 0.9609 &\vline& 0.6689 & 0.9681 &\vline& 
0.6695 & 0.9632 &\vline& 0.5740 & 0.9425 &\vline& 0.5805 & 0.9383 \\
4PN &\vline& 0.9174 & 0.9303 &\vline& 0.9087 & 0.9176 &\vline& 0.6693 & 0.7227 &\vline& 
0.6701 & 0.7230 &\vline& 0.6129 & 0.7112 &\vline& 0.6236 & 0.7159 \\
4.5PN &\vline& 0.9525 & 0.9744 &\vline& 0.9330 & 0.9415 &\vline& 0.7829 & 0.9242 &\vline& 
0.7827 & 0.9229 &\vline& 0.7286 & 0.9689 &\vline& 0.7308 & 0.9632 \\
5PN &\vline& 0.9370 & 0.9392 &\vline& 0.9225 & 0.9241 &\vline& 0.7275 & 0.7417 &\vline& 
0.7276 & 0.7420 &\vline& 0.6972 & 0.7409 &\vline& 0.7027 & 0.7500 \\
\hline
\hline
\label{table:Faithfulness-SC-WN}
\end{tabular}
\end {table}	

First, we explore the general question of the closeness of the standard adiabatic 
and complete adiabatic templates to the exact waveform assuming flat 
power spectral density for the detector noise. 
Figs.~\ref{fig:systemB-WN}--\ref{fig:systemE-WN} show the effectualness and 
faithfulness of various PN templates for three archetypical binaries
with component masses ($1M_{\odot},10M_{\odot}$), ($1M_{\odot},50M_{\odot}$) and  
($1M_{\odot},100M_{\odot}$), respectively~\footnote{For the sake of convenience 
we also tabulate the results shown in Figs.~\ref{fig:systemB-WN}--\ref{fig:systemE-WN} 
in Tables \ref{table:Effectualness-SC-WN} and \ref{table:Faithfulness-SC-WN}.}.

The central result of this study is that {\it complete adiabatic approximants 
bring about a remarkable improvement in the effectualness for all systems at 
low  PN orders ($<$ 3PN).} The complete adiabatic approximants  converge to the 
adiabatic exact waveform at lower PN orders than the standard adiabatic 
approximants. This indicates that at these orders general relativistic corrections 
to the conservative dynamics of the binary are quite important contrary to
the assumption employed in the standard post-Newtonian treatment of the
phasing formula. On the other hand, the difference in effectualness between 
the  standard and complete adiabatic approximants at  orders greater than 
3PN is very small. Thus, {\it if we have a sufficiently accurate 
(order $\geq$ 3PN) T-approximant of the flux function, the standard adiabatic 
approximation is nearly as good as the complete adiabatic approximation for
construction of effectual templates.} But at all orders the standard adiabatic 
approximation provides a lower bound to the complete adiabatic approximation 
for the construction of {\it effectual} templates. 

The faithfulness of both the approximants fluctuates as we go from one PN 
order to the next and is generally much smaller than our target value of 
0.965. The fluctuation continues all the way up to 5PN order reflecting the 
oscillatory approach of the flux function to the exact flux function at 
different PN orders. It is again interesting to note that complete adiabatic 
approximants are generally more faithful than the standard adiabatic approximants. 
It is certainly worth exploring, in a future study, the anomalous cases where 
it performs worse than the standard.

\begin{figure*}[t]
\centering \includegraphics[width=5in]{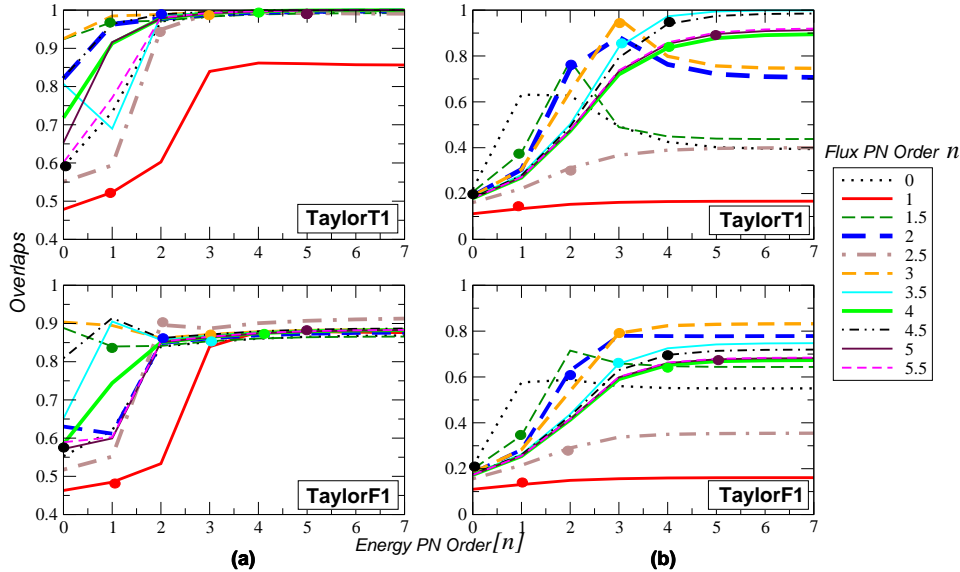}
\caption{Effectualness (left panels) and faithfulness (right panels) 
of various test mass {\it TaylorT1} and {\it TaylorF1} templates in 
detecting a signal from a $(1 M_\odot,10 M_\odot)$, calculated for 
the initial LIGO noise PSD.  Different lines in the panels correspond 
to different orders of the flux function. Each line shows how the 
overlaps are evolving as a function of the accuracy of the energy 
function. Standard adiabatic approximants $T(E_{[n]},{\cal F}_{n})$ are marked 
with thick dots.}
\label{fig:systemB-LIGO}
\end{figure*}

\subsubsection{ Initial LIGO noise spectrum}

\begin{figure*}[t]
\centering \includegraphics[width=5in]{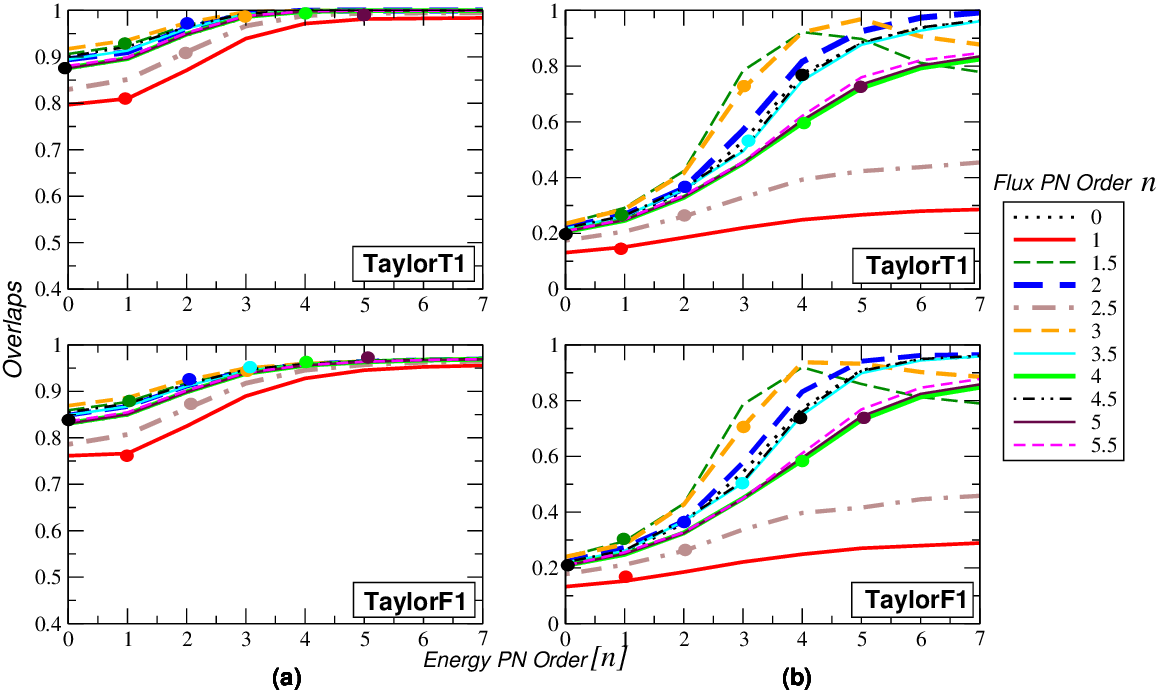}
\caption{As in Fig.~\ref{fig:systemB-LIGO} except that the signal 
corresponds to a $(1 M_\odot,50 M_\odot)$ binary.}
\label{fig:systemD-LIGO}
\end{figure*}

Having addressed the general question concerning the closeness of standard 
adiabatic and complete adiabatic templates to the exact waveforms, we now 
compare the overlaps in the specific case of the initial LIGO 
detector.
The effectualness and faithfulness of various test mass PN templates for the 
$(1M_{\odot},10 M_{\odot})$ binary and $(1M_{\odot},50 M_{\odot})$ binary 
are plotted in Fig.~\ref{fig:systemB-LIGO} and Fig.~\ref{fig:systemD-LIGO}, 
respectively, and are shown in Tables \ref{table:Effectualness1-SC-IL},  
\ref{table:Effectualness2-SC-IL} and \ref{table:Faithfulness-SC-IL}. 
As in the case of white-noise, here too we see  that standard adiabatic 
approximants of order less than 3PN have considerably lower overlaps than 
the corresponding complete adiabatic  approximants and the difference in 
overlaps between  standard adiabatic and complete adiabatic approximants of 
order $\geq$ 3PN is very small. Thus, as in the white-noise case, if we have 
a sufficiently accurate (order $\geq$ 3PN) T-approximant of the flux function, 
the standard adiabatic  approximation is nearly as good as the complete adiabatic 
approximation for the construction of effectual templates. Unlike in the 
white-noise case, for the initial LIGO noise spectrum the plots and 
Table~\ref{table:Faithfulness-SC-IL} indicate that the {\it faithfulness 
of PN templates greatly improves in a complete adiabatic treatment,  for all orders} 
studied.

We also calculate the bias in the estimation of parameters while maximizing 
the overlaps over the intrinsic parameters of the binary. The (percentage) bias 
in the estimation of the parameter $p$ is defined as
\begin{equation}
\sigma_{p} \equiv \frac{|p_{\rm max} - p|}{p} \times 100 \,,
\label{eq:bias}
\end{equation}
where $p_{\rm max}$ is the value of the parameter $p$ which gives the maximum 
overlap. Along with the maximized overlaps (effectualness), the bias in the 
estimation of the parameters $m$ and $\eta$ are also quoted in Tables 
\ref{table:Effectualness1-SC-IL},  \ref{table:Effectualness2-SC-IL}. 
It can be seen that at lower PN orders (order $<$ 3PN) the  complete adiabatic 
approximants show significantly lower biases. Even at higher PN orders complete 
adiabatic approximants are generally less-biased than the corresponding standard 
adiabatic approximants. 

\begin{table} 
\caption{Effectualness of {\it standard adiabatic} $T(E_{[n]},{\cal F}_{n})$ and {\it complete adiabatic}  
$T(E_{[n+2.5]},{\cal F}_{n})$ approximants in the test mass limit. Overlaps are calculated for the 
initial LIGO noise spectrum. Percentage biases $\sigma_{m}$ and $\sigma_{\eta}$ in determining 
parameters $m$ and $\eta$ are given in brackets.}
\begin{tabular}{ccccccccccccc}
\hline
\hline
&\vline&\multicolumn{7}{c}{$(1M_{\odot},10M_{\odot})$} \\
\cline{2-9}
&\vline&\multicolumn{3}{c}{{\it TaylorT1}}&\vline&\multicolumn{3}{c}{{\it TaylorF1}}\\
\cline{2-9}
Order ($n$)&\vline& {\it S} && {\it C} &\vline&  {\it S} && {\it C}\\ 
\hline
\hline

0PN   &\vline& 0.5910 (12, 5.7)  && 0.9707 (36, 45)   &\vline& 0.5527 (31, 28)  && 0.8395 (48, 53) \\
1PN   &\vline& 0.5232 (22, 105)  && 0.8397 (125, 69)  &\vline& 0.4847 (18, 9.7) && 0.8393 (147, 74) \\
1.5PN &\vline& 0.9688 (52, 51)   && 0.9887 (8.3, 15)  &\vline& 0.8398 (61, 57)  && 0.8606 (4.7, 10) \\
2PN   &\vline& 0.9781 (18, 25)   && 0.9942 (0.4, 0.6) &\vline& 0.8485 (32, 40)  && 0.8693 (15, 22) \\
2.5PN &\vline& 0.9490 (96, 68)   && 0.9923 (26, 32)   &\vline& 0.8963 (123, 75) && 0.9071 (49, 50) \\
3PN   &\vline& 0.9942 (0.3, 1.1) && 0.9989 (3.7, 6.2) &\vline& 0.8713 (16, 23)  && 0.8822 (12, 18) \\
3.5PN &\vline& 0.9940 (6.9, 11)  && 0.9998 (0.6, 1.4) &\vline& 0.8685 (23, 31)  && 0.8834 (17, 25) \\
4PN   &\vline& 0.9974 (6.2, 11)  && 0.9996 (3.9, 6.9) &\vline& 0.8746 (23, 30)  && 0.8817 (21, 28) \\
4.5PN &\vline& 0.9988 (3.3, 5.5) && 1.0000 (0.8, 1.6) &\vline& 0.8795 (19, 27)  && 0.8868 (18, 26) \\
5PN   &\vline& 0.9992 (4.0, 6.9) && 0.9997 (3.5, 5.7) &\vline& 0.8792 (21, 29)  && 0.8825 (20, 28) \\

\hline
\hline
\label{table:Effectualness1-SC-IL}
\end{tabular}
\end {table}

\begin{table} 
\caption{Same as Table ~\ref{table:Effectualness1-SC-IL} except that the values corresponds to the 
$(1M_{\odot},50M_{\odot})$ binary.}
\begin{tabular}{ccccccccccccc}
\hline
\hline
&\vline & \multicolumn{7}{c}{$(1M_{\odot},50M_{\odot})$}\\
\cline{2-9}
&\vline&\multicolumn{3}{c}{{\it TaylorT1}}&\vline&\multicolumn{3}{c}{{\it TaylorF1}}\\
\cline{2-9}
Order ($n$)&\vline& {\it S} && {\it C} &\vline&  {\it S} && {\it C}\\
\hline
\hline

0PN   &\vline& 0.8748 (24, 29)  && 0.9471 (19, 14)   &\vline& 0.8294 (21, 34)  && 0.8974 (17, 13)   \\
1PN   &\vline& 0.8101 (28, 104) && 0.9392 (19, 40)   &\vline& 0.7662 (23, 116) && 0.8898 (18, 43)   \\
1.5PN &\vline& 0.9254 (21, 4.1) && 0.9996 (6.7, 20)  &\vline& 0.8772 (18, 0.2) && 0.9590 (6.4, 20)  \\
2PN   &\vline& 0.9610 (18, 16)  && 0.9993 (7.5, 16)  &\vline& 0.9113 (16, 14)  && 0.9583 (7.7, 17)  \\
2.5PN &\vline& 0.9104 (21, 6.9) && 0.9940 (8.3, 0.7) &\vline& 0.8630 (19, 8.7) && 0.9574 (9.1, 1.9) \\
3PN   &\vline& 0.9968 (11, 21)  && 0.9992 (2.6, 10)  &\vline& 0.9500 (11, 21)  && 0.9648 (2.7, 11)  \\
3.5PN &\vline& 0.9923 (13, 19)  && 0.9997 (2.4, 5.2) &\vline& 0.9445 (12, 18)  && 0.9679 (2.8, 6.5) \\
4PN   &\vline& 0.9979 (8.8, 13) && 0.9995 (3.5, 4.3) &\vline& 0.9560 (8.9, 14) && 0.9672 (3.9, 5.6) \\
4.5PN &\vline& 0.9995 (7.1, 14) && 1.0000 (0.9, 1.9) &\vline& 0.9590 (7.0, 14) && 0.9698 (1.5, 3.4) \\
5PN   &\vline& 0.9994 (5.2, 7.7)&& 0.9990 (2.6, 2.4) &\vline& 0.9634 (5.9, 10) && 0.9690 (3.4, 5.1) \\

\hline
\hline
\label{table:Effectualness2-SC-IL}
\end{tabular}
\end {table}	

\begin{table} 
\caption{Faithfulness of {\it standard adiabatic} $T(E_{[n]},{\cal F}_{n})$ and {\it complete adiabatic}  
$T(E_{[n+2.5]},{\cal F}_{n})$ templates in the test mass limit. Overlaps are calculated for the 
initial LIGO noise spectrum.}
\begin{tabular}{ccccccccccccccccccc}
\hline
\hline
&\vline&\multicolumn{5}{c}{$(1M_{\odot},10M_{\odot})$} &\vline & \multicolumn{5}{c}{$(1M_{\odot},50M_{\odot})$}\\
\cline{2-13}
&\vline&\multicolumn{2}{c}{{\it TaylorT1}}&\vline&\multicolumn{2}{c}{{\it TaylorF1}}
&\vline&\multicolumn{2}{c}{{\it TaylorT1}}&\vline&\multicolumn{2}{c}{{\it TaylorF1}}\\
\cline{2-13}
Order ($n$)&\vline& {\it S} & {\it C} &\vline&  {\it S} & {\it C} 
&\vline& {\it S} & {\it C} &\vline&  {\it S} & {\it C}\\
\hline
\hline

0PN   &\vline& 0.2186 & 0.6272 &\vline& 0.2108 & 0.5879 &\vline& 0.2134 & 0.3498 &\vline& 0.2145 & 0.3593\\
1PN   &\vline& 0.1342 & 0.1615 &\vline& 0.1308 & 0.1563 &\vline& 0.1511 & 0.2196 &\vline& 0.1527 & 0.2210\\
1.5PN &\vline& 0.3788 & 0.4492 &\vline& 0.3449 & 0.6471 &\vline& 0.2915 & 0.9223 &\vline& 0.2956 & 0.9195\\
2PN   &\vline& 0.7449 & 0.7633 &\vline& 0.6279 & 0.7782 &\vline& 0.3613 & 0.8157 &\vline& 0.3674 & 0.8318\\
2.5PN &\vline& 0.3115 & 0.3970 &\vline& 0.2905 & 0.3532 &\vline& 0.2608 & 0.4233 &\vline& 0.2606 & 0.4161\\
3PN   &\vline& 0.9633 & 0.7566 &\vline& 0.7913 & 0.8297 &\vline& 0.7194 & 0.9686 &\vline& 0.7057 & 0.9323\\
3.5PN &\vline& 0.8385 & 0.9984 &\vline& 0.6582 & 0.7464 &\vline& 0.4941 & 0.9273 &\vline& 0.5046 & 0.9442\\
4PN   &\vline& 0.8356 & 0.8909 &\vline& 0.6527 & 0.6725 &\vline& 0.5960 & 0.7934 &\vline& 0.5864 & 0.8131\\
4.5PN &\vline& 0.9395 & 0.9851 &\vline& 0.6967 & 0.7195 &\vline& 0.7594 & 0.9644 &\vline& 0.7605 & 0.9614\\
5PN   &\vline& 0.8960 & 0.9129 &\vline& 0.6770 & 0.6821 &\vline& 0.7344 & 0.8350 &\vline& 0.7432 & 0.8579\\

\hline
\hline
\label{table:Faithfulness-SC-IL}
\end{tabular}
\end {table}	

\subsubsection{Accuracy of energy function Vs.~flux function}

In most of the cases, TaylorT1 and TaylorF1 templates show trends of 
smoothly increasing overlaps as the accuracy of the energy function 
is increased keeping the accuracy of the flux function constant. This 
is because the T-approximants of the energy function smoothly 
converge to the exact energy as we go to higher orders (see 
Fig.~\ref{fig:TMEnFlfnsLigo}$a$). On the other hand, if we improve 
the accuracy of the flux function for a fixed order of energy, the 
overlaps do not show such a smoothly converging behavior. This can be 
understood in terms of the oscillatory nature of the T-approximants 
of the flux function. For example, templates constructed from 1PN and 
2.5PN flux functions can be seen to have considerably lower overlaps 
than the other ones. This is because of the poor ability of the 1PN 
and 2.5PN T-approximants to mimic the behavior of the exact flux 
function (see Fig.~\ref{fig:TMEnFlfnsLigo}$b$). This  inadequacy
of the 1PN and 2.5PN T-approximants is prevalent in both test mass 
and comparable mass cases. Hence it is not a good  strategy to use the 
T-approximants at these orders for the construction of templates. On 
the other hand, 3.5PN and 4.5PN T-approximants are greatly successful 
in  following  the exact flux function in the test mass case, and  
consequently lead to  larger overlaps.

We have found that in the test mass case if we improve the accuracy 
of energy function from 3PN to 4PN, keeping the flux function at 
order 3PN, the increase in effectualness (respectively, faithfulness) 
is $\simeq 0.36\%$ $(-16\%)$. The same improvement in the energy 
function for the  3.5PN flux will produce an increase of $\simeq 
0.36\%$ $(13\%)$. On the other hand, if we improve the accuracy of flux 
function from 3.5PN to 4PN, keeping the energy function at order 3PN, 
the increase in effectualness (respectively, faithfulness) 
is $\simeq -0.17\%$ $(-12\%)$. The values quoted are calculated 
using the TaylorT1 method for the ($1M_{\odot},\, 10M_{\odot})$ binary 
for the initial LIGO noise PSD. The effectualness trends are 
similar in the case of the $(1M_{\odot},\, 50M_{\odot})$
binary also. {\it If the comparable mass case is qualitatively similar 
to the test mass case, this should imply that neither the improvement in 
the accuracy of the energy function from 3PN to 4PN nor the improvement 
in the accuracy of the flux function from 3.5PN to 4PN will produce 
significant improvement in the effectualness in the comparable mass case.}
The trends in the faithfulness are very different for different binaries 
so that it is hard to make any statement about the improvement
in faithfulness. 

\subsubsection{Number of gravitational wave cycles}

The number of GW cycles accumulated by a template is defined as~\cite{dis02}

\begin{equation}
{\cal N}_{\rm tot} \equiv \frac{1}{2 \pi}(\varphi_{\rm lso} - \varphi_{\rm low}) 
= \int_{F_{\rm low}}^{F{\rm lso}}dF \frac{N(F)}{F} \,,
\label{eq:numcycles}
\end{equation}
where $\varphi_{\rm lso}$ and $\varphi_{\rm low}$ are the GW phases corresponding 
to the last stable orbit and the low frequency cut-off, respectively, and 
$N(F) \equiv F^2/\dot{F}$ is the {\it instantaneous number of cycles}
spent near some instantaneous frequency $F$ (as usual, $\dot{F}$ is the time 
derivative of $F$). However, it has been noticed that \cite{dis02}, the large 
number ${\cal N}_{\rm tot}$ is not significant because the only really {\it useful} 
cycles are those that contribute most to the signal-to-noise ratio (SNR). The number 
of {\it useful cycles} is defined as  \cite{dis02}
\begin{equation}
{\cal N}_{\rm useful} \equiv \left(\int_{F_{\rm low}}^{F{\rm lso}}\frac{df}{f} w(f)N(f)\right) 
	\left(\int_{F_{\rm low}}^{F{\rm lso}}\frac{df}{f} w(f)\right)^{-1},
\label{eq:usecycles}
\end{equation}
where  $w(f) \equiv a^2(f)/h^2_n(f)$. If $S_n(f)$ is the two-sided PSD of the 
detector noise, $h_n(f)$ is defined by $h_n^2(f) \equiv fS_n(f)$, while $a(f)$ is 
defined by $|H(f)| \simeq a(t_f)/[{\dot F(t_f)}]^{1/2}$ where $H(f)$ is the Fourier 
transform of the time-domain waveform $h(t)$ (See Eqs.(\ref{eq:waveform1}) and 
(\ref{eq:waveform2})) and $t_f$ is the time when the instantaneous frequency $F(t)$ 
reaches the value $f$ of the Fourier variable.

The total numbers of GW cycles accumulated by various standard adiabatic 
$T(E_{[n]},{\cal F}_{n})$ and complete adiabatic 
$T(E_{[n+2.5]},{\cal F}_{n})$ approximants  in the test mass limit are
tabulated in Table~\ref{table:cycles20} along with the number of useful 
cycles calculated for the initial LIGO noise PSD. We use Eq.(\ref{eq:phasing2}) 
to calculate $\dot F$ and numerically evaluate the integrals in Eq.(\ref{eq:usecycles}) 
to compute the number of useful cycles. In order to compute the total number of cycles, 
we numerically evaluate the integral in Eq.(\ref{eq:numcycles})

It can be seen that all complete adiabatic approximants accumulate fewer
number of (total and useful) cycles than the corresponding standard adiabatic 
approximants. This is because the additional conservative terms in 
the complete adiabatic approximants add extra acceleration to the test mass 
which, in the presence of radiation reaction, would mean that the test body has to 
coalesce faster, and therefore such templates accumulate fewer number of cycles.
Notably enough, approximants (like 3PN and 4.5PN) producing the highest overlaps 
with the exact waveform, accumulate the closest number of cycles as 
accumulated by the exact waveform. This is indicative that the phase evolution 
of these approximants is closer to that of the exact waveform. On the other hand, 
the fractional absolute difference in the number of cycles of the approximants 
producing the lowest overlaps (like 0PN, 1PN and 2.5PN) as compared to the exact 
waveform is the greatest, which indicates that these templates follow a completely 
different phase evolution. 

\begin{figure}[t]
\centering \includegraphics[width=5.5in]{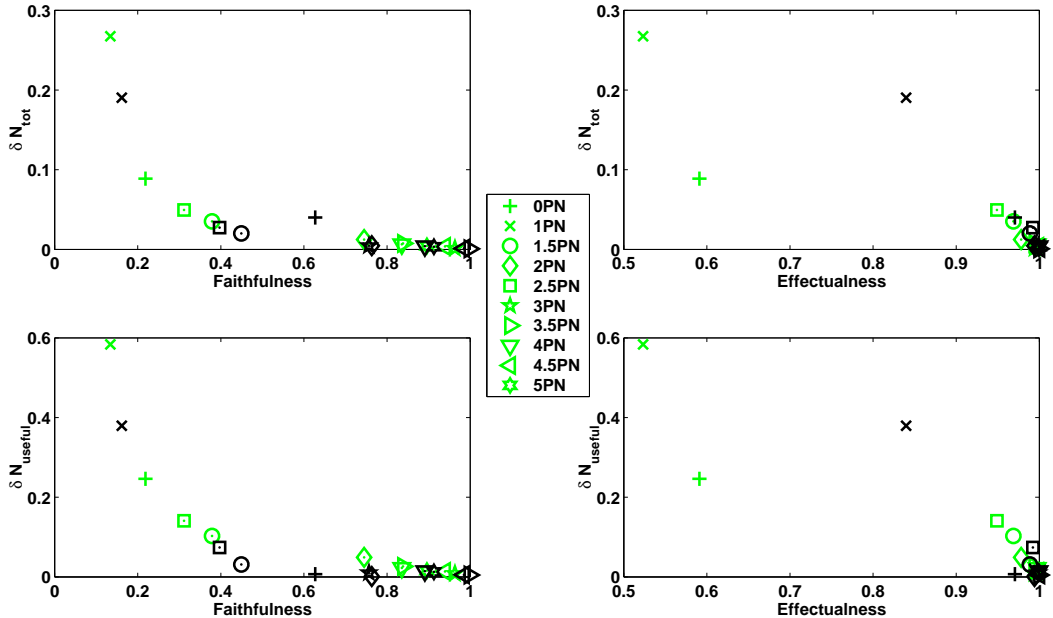}
\caption{The fractional absolute difference $\delta{\cal N}_{n}$ between 
the number of cycles accumulated by various approximants and the 
exact waveform, plotted against the corresponding overlaps. 
{\it Standard adiabatic} $T(E_{[n]},{\cal F}_{n})$ approximants 
are marked with lighter markers and {\it complete adiabatic} $T(E_{[n+2.5]},{\cal F}_{n})$ 
approximants are marked with darker markers. Top panels show $\delta{\cal N}_{n}$
for the total number of cycles and bottom panels show $\delta{\cal N}_{n}$
 for the number of useful cycles. The number of useful cycles are calculated for 
the initial LIGO noise PSD and the low frequency cut-off is chosen to be 40Hz. 
The plotted results are for a $(1M_{\odot},10M_{\odot})$ binary.}
\label{fig:cycles_overlap}
\end{figure}

In order to illustrate the correlation between the number of (total/useful) cycles accumulated
by an approximant and its overlap with the exact waveform, we introduce a quantity 
$\delta{\cal N}_{n} = \frac{|{\cal N}_{n} - {\cal N}_{\rm exact}|}{ {\cal N}_{\rm exact}}$
which is the fractional absolute difference between the number of (total/useful) cycles 
accumulated by a template and the exact waveform. Here ${\cal N}_{n}$ and  
${\cal N}_{\rm exact}$ are the  number of (total/useful) cycles accumulated by the $n$PN 
approximant and exact waveform, respectively. In Fig. \ref{fig:cycles_overlap}, we 
plot $\delta {\cal N}_{n}$ of various standard adiabatic and complete adiabatic approximants 
against the corresponding overlaps in the case of a $(1M_{\odot},10M_{\odot})$ binary.

The following points may be noted while comparing the results quoted here for the 
number of cycles with those of other works, e.g. Refs.~\cite{MSSTT,BFIJ02,AISS}.
As emphasized in Ref.~\cite{dis03} one can get very different results for the phasing 
depending on whether one consistently re-expands the constituent
energy and flux functions or evaluates them without re-expansion. In the computation 
of the number of useful cycles different authors treat the function ${\dot F}$
differently, some re-expand and others do not, leading to differences in the results.
The other important feature we would like to comment upon is a result that appears, 
at first, very counter-intuitive. It is the fact that in some cases the number of 
useful GW cycles is greater than the total number of GW cycles! A closer examination 
reveals that while for most cases of interest this does not happen, in principle its 
occurrence is determined by the ratio $f_{\rm r} \equiv F_{\rm low}/F_{\rm lso}$. 
To understand this in more detail let us consider the ratio ${\cal N}_{\rm r}$ 
of the number of useful cycles to the total number of cycles in the case of 
white-noise (in a frequency band $F_{\rm low}$ to $F_{\rm lso}$) for which 
\begin{equation}
{\cal N}_{\rm r} \equiv \frac{{\cal N}_{\rm useful}}{{\cal N}_{\rm tot}} = 
\frac{5}{12}f_{\rm r}^{1/3}\frac{1-f_{\rm r}^{4/3}}{1-f_{\rm r}^{1/3}}\,.
\end{equation}
For $f_{\rm r}\ll 1$, ${\cal N}_{\rm r} < 1$. However, as $f_{\rm r}$ increases to about 
$f_{\rm r}=0.52,$ ${\cal N}_{\rm r}$ transits from being less than one to becoming 
greater than one! Essentially this arises due to 
the details of the scalings of the various quantities involved and the point of 
transition depends on the PN order and the precise form of the noise PSD. 
For $f_{\rm r} \simeq 1$, the calculation of useful cycles does not make 
much physical sense. This explains the absence of ${\cal N}_{\rm useful}$ 
results for the $(1M_{\odot},100M_{\odot})$ binary in Table \ref{table:cycles20}. 

\begin{table} 
\caption{Number of GW cycles accumulated by various {\it standard adiabatic} 
$T(E_{[n]},{\cal F}_{n})$ and {\it complete adiabatic} $T(E_{[n+2.5]},{\cal F}_{n})$ 
approximants  in the test mass limit. The number of useful cycles calculated for the 
initial LIGO noise PSD is also shown in brackets. We choose a low frequency cut-off of 40Hz.}
\begin{tabular}{ccccccccccccccccccc}
\hline
\hline
&\vline&\multicolumn{3}{c}{$(1M_{\odot},10M_{\odot})$} &\vline & \multicolumn{3}{c}
{$(1M_{\odot},50M_{\odot})$} &\vline & \multicolumn{3}{c}{$(1M_{\odot},100M_{\odot})$} \\
\cline{2-13}
Order ($n$) &\vline&{\it S}&&{\it C}&\vline&{\it S}&&{\it C}&\vline&{\it S}&&{\it C}\\
\hline
\hline

0PN    &\vline& 481 (92.3) && 424 (74.6) &\vline& 118 (110)   && 77.8 (64.4) &\vline& 13.6 &&  6.7 \\ 
1PN    &\vline& 560 (117)  && 526 (102)  &\vline& 180 (186)   && 124  (104)  &\vline& 25.7 && 10.6 \\ 
1.5PN  &\vline& 457 (81.7) && 433 (71.8) &\vline& 88.8 (76.3) && 58.5 (38.2) &\vline&  8.4 &&  2.3 \\ 
2PN    &\vline& 447 (77.7) && 440 (74.0) &\vline& 77.0 (61.8) && 62.5 (41.5) &\vline&  6.1 &&  2.6 \\ 
2.5PN  &\vline& 464 (84.5) && 454 (79.6) &\vline& 96.8 (85.5) && 74.5 (50.5) &\vline&  9.7 &&  2.9 \\ 
3PN    &\vline& 442 (74.7) && 440 (73.3) &\vline& 64.5 (45.2) && 58.1 (35.5) &\vline&  3.4 &&  1.6 \\ 
3.5PN  &\vline& 445 (76.1) && 442 (74.5) &\vline& 68.7 (49.7) && 60.6 (36.8) &\vline&  4.0 &&  1.4 \\ 
4PN    &\vline& 445 (75.8) && 443 (75.2) &\vline& 66.4 (45.1) && 62.9 (39.0) &\vline&  2.9 &&  1.6 \\ 
4.5PN  &\vline& 443 (75.1) && 442 (74.5) &\vline& 63.7 (42.0) && 60.0 (35.6) &\vline&  2.5 &&  1.2 \\ 
5PN    &\vline& 444 (75.3) && 443 (75.0) &\vline& 63.8 (40.9) && 62.2 (37.8) &\vline&  2.1 &&  1.4 \\ 
Exact  &\vline& 442 (74.1) &&	  	 &\vline& 59.1 (34.3) &&    	     &\vline&  0.9 &&      \\ 

\hline
\hline
\label{table:cycles20}
\end{tabular}
\end {table}	

\section {Non-adiabatic  models }
\label{sec:non-adiabatic models}
Before introducing new  non-adiabatic  models in  this section,
let us recapitulate our  point of view  in summary.
Contrary to the  standard adiabatic approximant which is constructed
from consistent PN expansions of the energy and flux functions
to the same relative PN order, we considered a new complete adiabatic approximant 
(still based on PN expansions of the energy and flux functions but of different
PN orders) but consistent with a complete PN acceleration.
Viewed in terms of the acceleration terms they include,
the standard adiabatic approximation is inconsistent by neglect of 
some  intermediate  PN order terms in the acceleration.
The complete adiabatic approximation on the other hand is constructed to
consistently include all the  relevant PN acceleration terms
neglected  in the associated standard approximant. These models 
were  a  prelude to phasing models constructed from the
dynamical equations of motion considered in this section.
 However, we have worked solely within the adiabatic approximation.
It is then  pertinent to ask whether one can construct natural 
non-adiabatic extensions of both the standard and complete adiabatic 
approximants. And if so, how do their performance compare?
Indeed, the  work of Buonanno and Damour \cite{bd00} within the effective
one-body approach to the dynamics did find differences between the
adiabatic and non-adiabatic solutions. In this Section we 
investigate whether it is possible to introduce non-adiabatic
formulations of the standard and complete approximants considered
in the previous Section. 

The Lagrangian models  studied by Buonanno, Chen and Vallisneri \cite{BCV02}
seem to be the natural candidates for the purpose since they are
specified by the acceleration experienced by the binary system.
The Lagrangian models considered in Ref.~\cite{BCV02} can be thought of as
the {\it standard non-adiabatic} approximants, since, following standard
choices, they lead to gaps in the post-Newtonian expansion of the acceleration.
Generalizing these Lagrangian models so that there are no  missing
PN terms, or gaps, in the acceleration we can construct the {\it complete non-adiabatic}
approximants. With a non-adiabatic variant of the standard and complete
approximants we can  then compare their relative performances. However,
we will be limited in this investigation because of two reasons: 
Firstly, the Lagrangian models are available only up to 3.5PN order, 
and higher order PN accelerations are as yet unavailable. Secondly, the only  
{\it exact} waveform we have, has however been constructed only in the 
{\it adiabatic} approximation.   Even in the test mass limit, the exact 
waveform is not known beyond the adiabatic approximation. Due to lack of anything 
better, we continue to use the exact waveform in the adiabatic approximation 
to measure the effectualness and faithfulness of the non-adiabatic approximants.

Thus to obtain non-adiabatic approximants, the signal is constructed by integrating 
the equations of motion directly using a Lagrangian formalism.  The equations  are 
schematically written as:
\begin{equation}
\frac{d\mathbf{x}}{dt} = \mathbf{v},\ \ \ \ 
\frac{d\mathbf{v}}{dt} = \mathbf{a}
\label{eq:NonAdiabEqnOfMotion}
\end{equation}
For the complete non-adiabatic  model of order $n$, 
all terms in the PN expansion for acceleration are retained up to order $n+2.5$
without any gaps.
For the standard non-adiabatic models, on the other hand,
only terms in the acceleration
consistent with the treatment of standard phasing 
are retained in the acceleration, resulting in gaps in the acceleration
corresponding to  intermediate PN terms neglected in the treatment.
E.g. for $n=0$ the standard non-adiabatic approximation
includes only the $\mathbf{a}_{\rm N}$ and $\mathbf{a}_{\rm 2.5PN}$
while the complete non-adiabatic approximation would include
in addition the $\mathbf{a}_{\rm 1PN}$ and $\mathbf{a}_{\rm 2PN}$.
Given the current status of knowledge
of the two-body equations
of motion, we have only two complete non-adiabatic approximants,
at 0PN and 1PN retaining all acceleration terms up to 2.5PN
and 3.5PN, respectively. The associated 0PN (1PN) standard non-adiabatic
approximation retains acceleration terms corresponding to
0PN and 2.5PN (0PN, 1PN, 2.5PN and 3.5PN).

The explicit  terms for accelerations
for each order are given as follows~\cite{BCV02,BI03,MoraWill}
\begin{eqnarray}
\mathbf{a}_{\rm{N}} & =  & -\frac{M}{r^2}\mathbf{\hat{n}}, \\
\mathbf{a}_{\rm{1PN}} & =  & -\frac{M}{r^2}\left\{\mathbf{\hat{n}}\left[(1+3\eta)v^2
- 2(2+\eta)\frac{M}{r} - \frac{3}{2}\eta\dot{r}^2\right] - 2(2-\eta)\dot{r}\mathbf{v}\right\},\\
\mathbf{a}_{\rm{2PN}} & = &-\frac{M}{r^2} \left\{ \mathbf{\hat{n}} \left[\frac{3}{4}(12+29\eta)
\left(\frac{M}{r}\right)^2 + \eta(3-4\eta)v^4 + \frac{15}{8}\eta(1-3\eta)\dot{r}^4 \right.\right.\nonumber \\
& - & \left.\left. \frac{3}{2}\eta(3-4\eta)v^2\dot{r}^2 
- \frac{1}{2}\eta(13-4\eta)\frac{M}{r}v^2 -
 (2+25\eta+2\eta^2)\frac{M}{r}\dot{r}^2\right] \right.\nonumber\\
& - & \left. \frac{1}{2}\dot{r}\mathbf{v}\left[\eta(15+4\eta)v^2 - (4+41\eta+8\eta^2)\frac{M}{r} 
- 3\eta(3+2\eta)\dot{r}^2\right]\right\},\\
\mathbf{a}_{\rm{3PN}} &=& -\frac{M}{r^2}\left\{\mathbf{\hat{n}}\left[-\frac{35\dot{r}^6\eta}{16}
+ \frac{175\dot{r}^6\eta^2}{16} - \frac{175\dot{r}^6\eta^3}{16} + \frac{15\dot{r}^4\eta v^2}{2} 
- \frac{135\dot{r}^4\eta^2v^2}{4} + \frac{255\dot{r}^4\eta^3v^2}{8}\right.\right.\nonumber \\
&-& \frac{15\dot{r}^2\eta v^4}{2} + \frac{237\dot{r}^2\eta^2v^4}{8} - \frac{45\dot{r}^2\eta^3v^4}{2}
 + \frac{11\eta v^6}{4} - \frac{49\eta^2 v^6}{4} + 13\eta^3 v^6 \nonumber \\
 &+& \frac{M}{r}\left(79\dot{r}^4\eta - \frac{69\dot{r}^4\eta^2}{2} 
- 30\dot{r}^4\eta^3 - 121\dot{r}^2\eta v^2 + 16\dot{r}^2\eta^2 v^2 + 20\dot{r}^2\eta^3v^2 \right.\nonumber \\
&+& \left. \frac{75\eta v^4}{4} + 8\eta^2 v^4 - 10\eta^3v^4\right) 
+ \frac{M^2}{r^2}\left(\dot{r}^2 + \frac{22717\dot{r}^2\eta}{168} 
+ \frac{11\dot{r}^2\eta^2}{8} - 7\dot{r}^2\eta^3 \right. \nonumber \\
&+& \left. \frac{615\dot{r}^2\eta\pi^2}{64} - \frac{20827\eta v^2}{840} 
+ \eta^3v^2 - \frac{123\eta\pi^2v^2}{64} 
\right) \nonumber \\
&+& \left. \frac{M^3}{r^3}\left(-16 -\frac{1399\eta}{12}
 - \frac{71\eta^2}{2} + \frac{41\eta\pi^2}{16}\right) \right] 
+ \mathbf{v}\left[-\frac{45\dot{r}^5\eta}{8} + 15\dot{r}^5\eta^2 \right. \nonumber \\
&+& \left. \frac{15\dot{r}^5\eta^3}{4} 
+ 12\dot{r}^3\eta v^2 - \frac{111\dot{r}^3\eta^2 v^2}{4} - 12\dot{r}^3\eta^3 v^2 
- \frac{65\dot{r}\eta v^4}{8} + 19\dot{r}\eta^2 v^4 + 6\dot{r}\eta^3v^4 \right. \nonumber \\ 
&+& \frac{M}{r}\left(\frac{329\dot{r}^3\eta}{6} 
+ \frac{59\dot{r}^3v^2}{2} + 18\dot{r}^3\eta^3 - 15\dot{r}\eta v^2
 - 27\dot{r}\eta^2 v^2 - 10\dot{r}\eta^3 v^2\right) \nonumber \\
 &+& \left. \left.\frac{M^2}{r^2}\left(-4\dot{r} - \frac{5849\dot{r}\eta}{840} + 25\dot{r}\eta^2 + 8\dot{r}\eta^3
 - \frac{123\dot{r}\eta\pi^2}{32} 
 \right)\right]\right\},
\label{eq:NonAdiabAccln3pn}\\
\mathbf{a}_{\rm{2.5RR}} & = & \frac{8}{5}\eta\frac{M^2}{r^3}\left\{ \dot{r}\mathbf{\hat{n}}\left[18v^2
+ \frac{2}{3}\frac{M}{r} - 25\dot{r}^2\right]
-\mathbf{v}\left[6v^2 - 2\frac{M}{r} - 15\dot{r}^2\right]\right\},\\
\mathbf{a}_{\rm{3.5RR}} &=& \frac{8}{5}\eta\frac{M^2}{r^3}\left\{ \dot{r}\mathbf{\hat{n}}
\left[\left(\frac{87}{14} -48\eta\right)v^4 - \left(\frac{5379}{28} + \frac{136}{3}\eta\right)v^2\frac{M}{r}
+\frac{25}{2}(1+5\eta)v^2\dot{r}^2 \right.\right.\nonumber \\
&+& \left.\left(\frac{1353}{4}+133\eta\right)\dot{r}^2\frac{M}{r}
- \frac{35}{2}(1-\eta)\dot{r}^4
+\left(\frac{160}{7} + \frac{55}{3}\eta\right)\left(\frac{M}{r}\right)^2\right]\nonumber \\
&-&\mathbf{v}\left[-\frac{27}{14}v^4 - \left(\frac{4861}{84}+\frac{58}{3}\eta\right)v^2\frac{M}{r} + \frac{3}{2}(13-37\eta)v^2\dot{r}^2 
\right. \nonumber \\
&+& \left.\left. \left(\frac{2591}{12}+97\eta\right)\dot{r}^2\frac{M}{r} - \frac{25}{2}(1-7\eta)\dot{r}^4 
+ \frac{1}{3}\left(\frac{776}{7}+55\eta\right)\left(\frac{M}{r}\right)^2\right]\right\},
\label{eq:NonAdiabAccln35pn}
\end{eqnarray}
where $\mathbf{\hat{n}} = \mathbf{r}/r$.
In the above Eq.(\ref{eq:NonAdiabAccln3pn}), the logarithmic terms  present in  $\mathbf{a}_{\rm 3PN}$ in 
\cite{BI03} have been transformed away by an infinitesimal gauge transformation 
following \cite{MoraWill} \footnote{We thank Luc Blanchet for pointing this to us and
providing us this form.}.

The above equations are solved numerically to attain the dynamics of the system. 
Then the orbital phase $\phi (t)$ is calculated by numerically solving the equations
\begin{equation} 
\frac{d \phi}{dt} = \omega, \,\,\,\,\,\, v^2 = \omega^2 r^2.
\end{equation}
where the calculation of the orbital angular frequency $\omega$ assumes that the orbit 
is circular. Once we have the orbital phase, the waveform is generated using 
Eq.(\ref{eq:waveform1}) since the orbital phase $\phi$ is related to the GW phase 
$\varphi$ by $\varphi = 2 \phi$.

\subsection{ Standard and complete non-adiabatic approximants in the
test mass  case}
\label{sec:non-adiab-tm-results} In the following we discuss the results 
of our study for the non-adiabatic waveforms in the test mass limit.
To determine the appropriate expressions for the acceleration
in the test mass case, we start with the general expression
for the acceleration and set $\eta=0$ in the conservative 1PN,
2PN and 3PN terms. Since doing the same in the dissipative terms
at 2.5PN and 3.5PN orders prevents the orbit from decaying, 
we retain terms linear in $\eta$ at these orders and set to zero terms of 
higher order in $\eta$. In the first part we discuss results found on
the problem of the closeness of the standard and complete non-adiabatic 
template families with  the  fiducial  exact waveform. In the second part 
we extend our results to the noise spectrum expected in initial LIGO.

\subsubsection{White noise}
\label{sec:non-adiab-tm-results-wn}
\begin{table}
\caption{Effectualness of the Lagrangian templates in the test mass case for the white-noise.}
\begin{tabular}{cccccccccc}
\hline
\hline
& \vline &\multicolumn{2}{c}{$(1M_{\odot}, 10M_{\odot})$}
& \vline &\multicolumn{2}{c}{$(1M_{\odot}, 50M_{\odot})$}
& \vline &\multicolumn{2}{c}{$(1M_{\odot},100M_{\odot})$}\\
\cline{2-10}
Order ($n$)
& \vline &  {\it S}  & {\it C}& \vline &  {\it S}  & {\it C}  & \vline & {\it S}  & {\it C} \\
\hline
\hline
0PN & \vline & 0.5521 & 0.4985 & \vline & 0.5553 & 0.8399 & \vline & 0.6775 & 0.7789 \\
1PN & \vline & 0.4415 & 0.4702 & \vline & 0.5760 & 0.8327 & \vline & 0.6557 & 0.7591 \\
\hline
\hline
\label{table:Effectualness-LM-TM-WN}
\end{tabular}
\end {table}

\begin{table}
\caption{Faithfulness of the Lagrangian templates in the test mass case for the white-noise.}
\begin{tabular}{cccccccccc}
\hline
\hline
& \vline &\multicolumn{2}{c}{$(1M_{\odot}, 10M_{\odot})$}
& \vline &\multicolumn{2}{c}{$(1M_{\odot}, 50M_{\odot})$}
& \vline &\multicolumn{2}{c}{$(1M_{\odot},100M_{\odot})$}\\
\cline{2-10}
Order ($n$)
& \vline &  {\it S}  & {\it C}& \vline &  {\it S}  & {\it C}  & \vline & {\it S}  & {\it C} \\
\hline
\hline
0PN & \vline & 0.0450 & 0.0441 & \vline & 0.1778 & 0.0991 & \vline & 0.5959 & 0.1851\\
1PN & \vline & 0.0471 & 0.0474 & \vline & 0.3195 & 0.1235 & \vline & 0.3646 & 0.2404\\
\hline
\hline
\label{table:Faithfulness-LM-TM-WN}
\end{tabular}
\end {table}

First, we explore the general question as to the closeness of the standard non-adiabatic 
and complete non-adiabatic templates assuming a flat power spectral density for the detector 
noise. Tables~\ref{table:Effectualness-LM-TM-WN} and \ref{table:Faithfulness-LM-TM-WN} 
show the effectualness and faithfulness of Lagrangian models for the same three archetypical
binaries as before: $(1M_{\odot},\, 10M_{\odot})$, $(1M_{\odot},\, 50M_{\odot})$ 
and $(1M_{\odot},\, 100M_{\odot})$ binaries. At present, the results are available at 
too few PN orders to make statements of general trends in effectualness and
faithfulness. However, the main result obtained for the adiabatic approximants seems to 
hold good again for the non-adiabatic approximants: the effectualness is higher for the 
complete non-adiabatic model as opposed to the standard non-adiabatic model. This is 
indicative of the fact that corrections coming from the conservative part of the dynamics 
(i.e. the well-known general relativistic effects at 1PN and 2PN) make an improvement of 
the effectualness. However, as in the adiabatic case, the faithfulness of both standard 
and complete non-adiabatic models is very poor. But, in sharp contrast to the adiabatic 
case, here it appears that the complete non-adiabatic approximation results in a decrease 
in the faithfulness of the templates. 

\subsubsection{Initial LIGO noise spectrum}
Having addressed the question concerning the closeness of standard non-adiabatic and 
complete non-adiabatic templates to exact waveforms, we now compare the overlaps in 
the case of the initial LIGO detector.

\begin{table}
\caption{Effectualness of the Lagrangian templates in the test mass 
case for the initial LIGO noise PSD. Percentage biases $\sigma_{m}$ 
and $\sigma_{\eta}$ in determining parameters $m$ and $\eta$ are given 
in brackets.}
\begin{tabular}{cccccccccc}
\hline
\hline
&\vline &\multicolumn{2}{c}{$(1M_{\odot}, 10M_{\odot})$}
&\vline &\multicolumn{2}{c}{$(1M_{\odot}, 50M_{\odot})$} \\
\cline{2-7}
Order ($n$) & \vline & {\it S} & {\it C} & \vline &  {\it S} & {\it C} \\
\hline
\hline
0PN    &\vline& 0.5848 (30, 26) & 0.9496 (55, 107) &\vline& 0.8741 (3.3, 9.2) & 0.9835 (35, 4.2) \\
1PN    &\vline& 0.6762 (37, 49) & 0.9273 (3.1, 27) &\vline& 0.8530 (34, 191)  & 0.9784 (24, 28) \\
\hline
\hline
\label{LagrangianEffectualnessLIGOTest}
\end{tabular}
\end{table}

\begin{table}
\caption{Faithfulness of the Lagrangian templates in the test mass case for 
the initial LIGO noise PSD.}
\begin{tabular}{cccccccccc}
\hline
\hline
&\vline &\multicolumn{2}{c}{$(1M_{\odot}, 10M_{\odot})$}
&\vline &\multicolumn{2}{c}{$(1M_{\odot}, 50M_{\odot})$}\\
\cline{2-7}
Order ($n$) & \vline & {\it S} & {\it C} & \vline &  {\it S} & {\it C}\\
\hline
\hline
0PN    & \vline & 0.2463 &0.1216 & \vline & 0.5048 & 0.1747 \\
1PN    & \vline & 0.4393 &0.1823 & \vline & 0.3650 & 0.3119 \\
\hline
\hline
\label{LagrangianFaithfulnessLIGOTest}
\end{tabular}
\end {table}

Tables~\ref{LagrangianEffectualnessLIGOTest} and \ref{LagrangianFaithfulnessLIGOTest} 
show the effectualness and faithfulness, respectively, of Lagrangian templates for the 
$(1M_{\odot},\, 10M_{\odot})$ and $(1M_{\odot},\, 50M_{\odot})$ binaries. In this 
case, we see the that the effectualness of the approximants gets significantly 
improved in the complete non-adiabatic approximation and is greater than 0.9 for 
all the  systems studied in this paper. Faithfulness appears to be decreased by the 
use of complete non-adiabatic approximation (this result also is in sharp contrast 
with the corresponding adiabatic case where we find that the complete approximation 
brings about a significant improvement in the faithfulness), but again there is no 
indication that either standard or complete templates are reliable in extracting 
the parameters of the system. 

\section {Comparable Mass Waveforms}
\label{sec:compmass}

In the case of comparable mass binaries there is no {\it exact} template 
available and the best we can do is to compare the performance of the standard 
adiabatic and complete adiabatic templates by studying their overlaps with 
some plausible fiducial `exact' waveform. As in the case of the test masses, 
here too we will consider all possible combinations of the T-approximants of the energy 
and flux functions, construct PN templates and calculate the overlaps of these 
templates with the fiducial `exact' waveform. In all cases, the fiducial 
`exact' waveform is constructed by numerically integrating the phasing formula 
in the time-domain (TaylorT1 approximant), and terminating the waveforms 
(`exact' and approximate) at $v_{lso}= 1/\sqrt{6}$ which corresponds to 
$F_{\rm lso}\simeq 1570$ Hz for a $(1.4M_{\odot},\, 1.4M_{\odot})$ binary and 
$F_{\rm lso}\simeq 220$ Hz for a $(10M_{\odot},\, 10M_{\odot})$ binary 
\footnote{Here also, $v_{\rm lso}$ is the velocity at the last stable circular orbit 
of the Schwarzschild geometry having the same mass as the total mass $m_1+m_2$ of 
the binary. Strictly speaking, in the comparable mass case, $v_{\rm lso}$ at 
$n$PN order should be determined by solving $E'_n(v)=0$ where $E'_n(v)$ is the 
$v$-derivative of the $n$th PN order energy function. Since we found that our
results are qualitatively independent of such considerations, we stick to the 
choice in the test-mass limit.}. The lower frequency cut-off of the waveforms
is chosen to be $F_{\rm low}= 40$ Hz. 

\subsection {The energy function}

Unlike in the test mass limit, the energy function $E(x;\eta)$ is not 
known exactly in the comparable mass case but only a 
post-Newtonian expansion, which has been computed at present up 
to 3PN accuracy \cite{DJS,BF,DJS02,BDE03,IF}. 
\begin{eqnarray}
E_{3PN} (x;\eta) & = & -\frac{1}{2} \, \eta \, x \, \left[ 1-\frac{1}{12} 
\right.\,
(9+\eta) \, x -\frac{1}{8} \, \left( 27 - 19\eta + \frac{\eta^2}{3} 
\right) \, x^2 \nonumber\\
 & + & \left(\frac{-\,675}{64} + \left( \frac{209323}{4032} - 
\frac{205 \pi^2}{96}\right. \right.
\left. - \frac {110\lambda}{9}\right)\eta- \frac{155}{96}\eta^2 
\left. - \frac {35}{5184}\eta^3 \right) \, x^3 
+ {\cal O} \,(x^4) \,\bigg]\,,
\label{eq:n18}
\end{eqnarray}
where $\lambda=-1987/3080 \simeq -0.6451$~\cite{DJS02,BDE03,IF,itoh2}. The corresponding $E'(v;\eta)$ 
appearing in the phasing formula reads,
\begin{eqnarray}
E'_{3PN}(v;\eta) & = & -\eta \, v \, \left[ 1-\frac{1}{6} \,\right.\,
(9+\eta) \, v^2 -\frac{3}{8} \, \left( 27 - 19\eta + \frac{\eta^2}{3} 
\right) \, v^4 \nonumber\\
& + & 4\left(\frac{-\,675}{64} + \left( \frac{209323}{4032} - 
\frac{205 \pi^2}{96} \right.\right.
\left. - \frac {110\lambda}{9}\right)\eta- \frac{155}{96}\eta^2 
\left.- \frac {35}{5184}\eta^3 \right) \, v^6 + {\cal O}(v^8) \,\bigg].
\label{eq:E_3pncompmass}
\end{eqnarray}

We use this expression truncated at the necessary orders to construct 
the various approximate templates. To compute a fiducial 
`exact' waveform, we use the exact energy function in the test mass 
limit supplemented by the finite mass corrections up to 3PN
in the spirit of the hybrid approximation \cite{KWW}.
In other words, the fiducial `exact' energy $E'(v;\eta)$ will look like 
\begin{eqnarray}
E'_{\rm exact}(v;\eta) & = & -\eta\,v\,\left[\frac{-\,E'_{\rm 
exact}(v)}{\eta\,v}-\frac{\eta}{6}\,v^2\right.\,
 -\frac{3}{8} \, \left(- 19 \,\eta + \frac{\eta^2}{3} \right) \, v^4 
\nonumber\\
& + & 4\left(\left( \frac{209323}{4032} - \frac{205 \, \pi^2}{96} 
\right.\right.
\left. - \frac {110\lambda}{9}\right)\eta- \frac{155}{96}\eta^2 
\left.\left.- \frac {35}{5184}\eta^3 \right) \, v^6 \, \right]
\label{eq:E_7pncompmass}
\end{eqnarray}
where $E'_{\rm exact}(v)$ is the v-derivative of the exact energy function in 
the test mass limit given by Eq.~(\ref{eq:Eprimeexact}). The T-approximants of 
the energy function $E'_{\rm T}(v;\eta)$ as well as the fiducial exact energy $E'_{\rm 
exact}(v;\eta)$ are plotted in Fig.~\ref{fig:CMEnFlFnLigo}$a$. The $v_{\rm lso}$ 
corresponding to the fiducial `exact' energy function can be determined by solving 
$E'_{\rm exact}(v;\eta) = 0$. 
This will yield a value $v_{\rm lso}^{\rm 3PN-hybrid}\simeq 0.4294$ against
the $v_{\rm lso}\simeq 0.4082$ in the test-mass case (more precisely it is the 
$v_{\rm MECO}$ \cite{BCV02}). If the $\eta$-corrections are included only up to 2PN instead 
of 3PN, $v_{\rm lso}^{\rm 2PN-hybrid}\simeq 0.4113$. It is worth pointing that 
$v_{\rm lso}^{\rm 2PN-Pade} \simeq 0.4456$ \cite{dis01} 
and it is not unreasonable to expect that, with 
3PN $\eta$-corrections the differences between various different ways of determining 
the lso converge. (For the purposes of our analysis, we have checked that there is no 
drastic change in our conclusions due to these differences and hence we use uniformly 
the value $v_{\rm lso}=0.4082$).

\subsection {The flux function}

\begin{figure}[t]
\centering \includegraphics[width=5in]{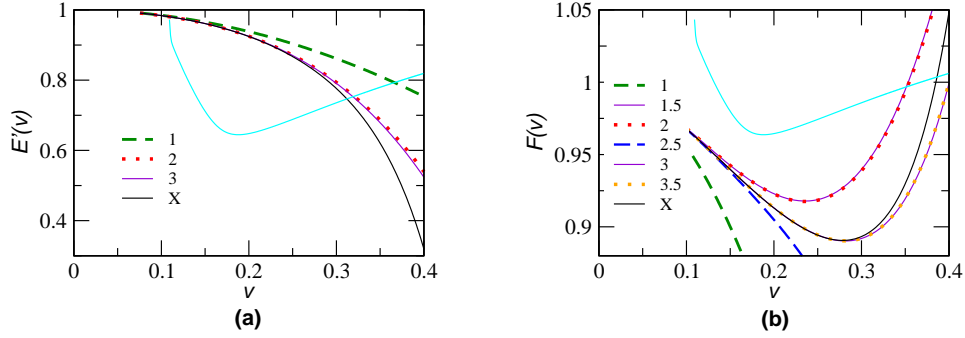}
\caption{Various T-approximants of Newton-normalized ($v$-derivative 
of) energy function $E'_{\rm T}(v)/E'_{\rm N}(v)$ (left) and flux 
function ${\cal F}_{\rm T}(v)/{\cal F}_{\rm N}(v)$ (right) in the 
comparable mass case, along with the corresponding fiducial `exact' 
functions (denoted by $X$). Also plotted is the amplitude spectral 
density (per $\sqrt{\rm Hz}$) of initial LIGO noise in arbitrary 
units.}
\label{fig:CMEnFlFnLigo}
\end{figure}

The flux function in the case of comparable masses has been 
calculated up to 3.5PN accuracy \cite{BDIWW,BDI,WW,BIWW,B96,BIJ02,BFIJ02}, 
and is given by:
\begin{eqnarray}
&&{\cal F}(v;\eta) =\frac{32}{5} \eta^2 v^{10}\left[ 
\sum_{k=0}^{7}A_k(\eta) v^k +B_6(\eta) v^6 \ln v + {\cal O} \,(v^8) \, 
\right],
\label{eq:3pnfluxcompmass}
\end{eqnarray}
where 
\allowdisplaybreaks{
\begin{eqnarray}
A_0 (\eta) & =   & 1, \ \ 
A_1 (\eta)   =    0, \ \ 
A_2 (\eta)   =    -\frac{1247}{336} - \frac{35}{12} \, \eta, 
\ \ 
A_3 (\eta)   =     4\pi, \nonumber \\
A_4 (\eta) & =  & -\frac{44711}{9072} + \frac{9271}{504} \, \eta +
\frac{65}{18} \, \eta^2, \\ 
A_5 (\eta) & = & -\left( \frac{8191}{672} + \frac{583}{24} \, \eta 
\right)\, \pi, \nonumber \\
A_6 (\eta) & =  & \frac{6643739519}{69854400} + \frac{16 \pi^2}{3} - 
\frac{1712}{105}\gamma  + \left( -\frac{11497453}{272160} + \frac{41 
\pi^2}{48} + \frac{176 \lambda}{9} - \frac{88 \Theta}{3}\right) \eta 
\nonumber \\
	   & - & \frac{94403}{3024}\eta^2 - \frac{775}{324}\eta^3 - 
\frac{1712}{105}\,{\rm ln}\, 4, \nonumber \\
A_7 (\eta) & =  & \left(-\frac{16285}{504} + \frac{214745}{1728}\eta 
+ \frac{193385}{3024}\eta^2\right) \,\pi, \\ 
B_6 (\eta) & = & - \frac{1712}{105},
\label{eq:fluxcoeffscompmass}
\end{eqnarray}}
and the value of $\Theta$ has been recently calculated to be 
$ -11831/9240 \simeq -1.28$ \cite{BDEI} by dimensional regularization. 

\begin{figure*}[t]
\centering \includegraphics[width=6.5in]{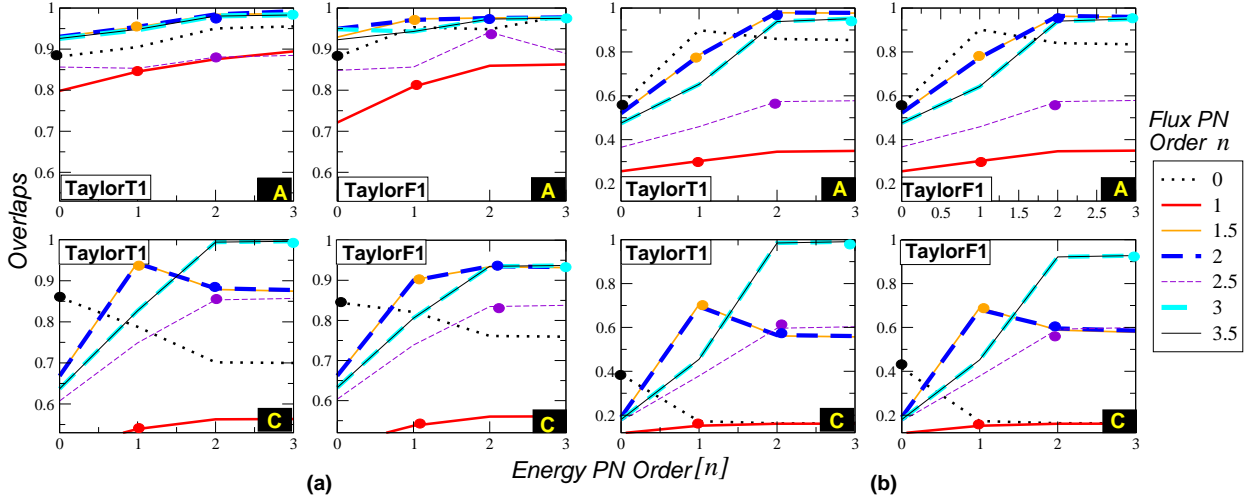}
\caption{Effectualness (Fig.~$a$) and Faithfulness (Fig.~$b$) of 
various TaylorT1 and TaylorF1 templates in the comparable 
mass case for the initial LIGO noise spectrum. Different lines in the 
panels correspond to different orders of the flux function. Each line 
shows how the overlaps are evolving as a function of the accuracy of the 
energy function.  Standard adiabatic approximants $T(E_{[n]},{\cal F}_n)$ 
are marked with thick dots. Label {\bf A} represents the 
$(10 M_\odot,\, 10 M_\odot)$ binary and label {\bf C} represents the 
$(1.4 M_\odot,\, 1.4 M_\odot)$ binary.}
\label{fig:4graphsCM}
\end{figure*}

To construct our fiducial 
`exact' waveform, we will use the energy function given by 
Eq.~(\ref{eq:E_7pncompmass}) and the flux function 
\begin{eqnarray}
{\cal F}_{\rm exact}(v;\eta) & = &\frac{32}{5} \eta^2 v^{10}\left[{\cal F}_{\rm 
exact}(v)-\sum_{k=0}^{7}\left(A_k v^k \right.\right.
\left.+  B_6 v^6 \ln v\right) 
+ \left.\sum_{k=0}^{7}\left(A_k(\eta) v^k +B_6(\eta) v^6 \ln v 
\right) \right],
\label{eq:55pnfluxcompmass}
\end{eqnarray}
where ${\cal F}_{\rm exact}(v)$ is the Newton-normalized 
(numerical) exact flux in the test-mass limit. The expansion 
coefficients $A_k$'s and $B_6$ refer to the test-mass 
case and $A_k(\eta)$'s and $B_6(\eta)$ refer to the comparable mass 
case.  The exact flux function is thus constructed by superposing
all that we know in the test mass case from perturbation
methods and the two body case by post-Newtonian methods.
It supplements the exact flux function in the test body limit by
all the $\eta$-dependent corrections known up to 3.5PN order in the
comparable mass case.  The T-approximants of the flux function 
${\cal F}_{\rm T}(v;\eta)$ and the fiducial exact flux 
${\cal F}_{\rm exact }(v;\eta)$ are plotted in Fig.~\ref{fig:CMEnFlFnLigo}$b$.

\subsection{Comparable mass results in the adiabatic approximation}

The effectualness and faithfulness of various PN templates in the case 
of comparable mass binaries are plotted in 
Fig.~\ref{fig:4graphsCM}$a$ and Fig.~\ref{fig:4graphsCM}$b$,  
respectively, and are tabulated in Tables 
\ref{table:Effectualness1-SC-CM-IL}, \ref{table:Effectualness2-SC-CM-IL}
and \ref{table:Faithfulness-SC-CM-WN}. The overlaps of the fiducial 
`exact' waveform are calculated with the TaylorT1 and TaylorF1 
approximants using the initial LIGO noise spectrum. Different 
lines in the panels of Fig.~\ref{fig:4graphsCM} correspond to 
different PN orders of the flux function. Let us note that in 
the case of comparable mass binaries the complete adiabatic 
approximants can be calculated, at present, at most up to 1PN order.
From the Tables \ref{table:Effectualness1-SC-CM-IL}, 
\ref{table:Effectualness2-SC-CM-IL} and \ref{table:Faithfulness-SC-CM-WN}
one can see that the complete adiabatic approximation generally improves 
the {\it effectualness} of the templates at 0PN and 1PN orders. 
But, as far as {\it faithfulness} is concerned, it is hard to 
conclude that one approximation is better than the other at these 
PN orders.

Even though complete adiabatic approximants are not calculated 
for higher PN orders, the general conclusion one can make from 
Fig.~\ref{fig:4graphsCM} is that the complete adiabatic approximation of the 
phasing will not result in a significant improvement in overlaps if 
we have a flux function of order $\geq$ 1.5PN. We, thus, conclude 
that, provided we have a sufficiently accurate (order $\geq$ 1.5PN) 
T-approximant of the flux function, the standard adiabatic 
approximation provides a good lower bound to the complete adiabatic 
approximation for the construction of both effectual and faithful 
templates in the case of comparable mass binaries. It should be kept in 
mind that unlike the test mass case where the exact energy and flux 
functions are known leading to an exact waveform in the adiabatic approximation, 
in the comparable mass  case we are only talking about {\it fiducial}
energy and flux functions constructed from what is known. Probably, the
fiducial waveform in this case has much less to do with the exact  waveform
predicted by general relativity.

Tables \ref{table:Effectualness1-SC-CM-IL} and \ref{table:Effectualness2-SC-CM-IL}
indicate that, to achieve the target sensitivity of 0.965 in effectualness corresponding 
to a 10\% loss in the event-rate, standard adiabatic approximants of order 2PN and 3PN 
are required for the $(10 M_\odot,10 M_\odot)$ and $(1.4 M_\odot,1.4 M_\odot)$ 
binaries, respectively, when restricting to only the inspiral phase. 
Even though the 2PN standard adiabatic approximants produce the required overlaps
in the case of the $(10 M_\odot,10 M_\odot)$ binary, in the real physical case 
of BH-BH binaries the inspiral family would not be adequate and must be supplemented 
by the plunge part of waveform as first discussed in \cite{bd00,dis03} and later 
in \cite{BCV02,DIJS}. A discussion of plunge requires a 3PN description of dynamics 
so that the 2PN templates are no longer adequate. 

\begin{table} 
\caption{Effectualness of {\it standard adiabatic} $T(E_{[n]},{\cal F}_{n})$ and {\it complete adiabatic}  
$T(E_{[n+2.5]},{\cal F}_{n})$ approximants in the comparable-mass case. Overlaps are calculated for the 
initial LIGO noise spectrum. Percentage biases $\sigma_{m}$ and $\sigma_{\eta}$ in determining 
parameters $m$ and $\eta$ are given in brackets.}
\begin{tabular}{ccccccccccccccc}
\hline
\hline
&\vline&\multicolumn{7}{c}{$(10M_{\odot},10M_{\odot})$} \\
\cline{2-9}
&\vline&\multicolumn{3}{c}{{\it TaylorT1}}&\vline&\multicolumn{3}{c}{{\it TaylorF1}}\\
\cline{2-9}
Order ($n$)&\vline& {\it S} && {\it C} &\vline&  {\it S} && {\it C}\\ 
\hline
\hline

0PN   	&\vline& 0.8815 (14, 0.2)  && 0.9515 (3.7, 0.1) &\vline& 0.8810 (11, 0.1)  && 0.9486 (4.1, 0.2)	\\
1PN   	&\vline& 0.8457 (59, 0.1)  && 0.8957 (45, 12)   &\vline& 0.8089 (52, 0.1)  && 0.8671 (40, 8.9)	\\
1.5PN 	&\vline& 0.9536 (3.9, 0.3) && 		        &\vline& 0.9738 (23, 29)   && 	              	\\
2PN   	&\vline& 0.9833 (0.4, 0.2) && 		        &\vline& 0.9762 (0.0, 1.9) && 		      	\\
2.5PN 	&\vline& 0.8728 (14, 0.1)  && 		        &\vline& 0.8800 (9.6, 0.1) && 			\\
3PN   	&\vline& 0.9822 (1.5, 0.0) && 		        &\vline& 0.9738 (1.3, 0.1) && 			\\
3.5PN 	&\vline& 0.9843 (1.4, 0.0) && 		        &\vline& 0.9746 (1.1, 0.1) && 		 	\\

\hline
\hline

\label{table:Effectualness1-SC-CM-IL}
\end{tabular}
\end {table}
	
\begin{table} 
\caption{Same as Table ~\ref{table:Effectualness1-SC-CM-IL} except that the values correspond to the 
$(1.4M_{\odot},1.4M_{\odot})$ binary.}
\begin{tabular}{ccccccccccccccc}
\hline
\hline
&\vline & \multicolumn{7}{c}{$(1.4M_{\odot},1.4M_{\odot})$}\\
\cline{2-9}
&\vline&\multicolumn{3}{c}{{\it TaylorT1}}&\vline&\multicolumn{3}{c}{{\it TaylorF1}}\\
\cline{2-9}
Order ($n$)&\vline& {\it S} && {\it C} &\vline&  {\it S} && {\it C}\\
\hline
\hline

0PN   	&\vline& 0.8636 (1.4, 0.2) && 0.6993 (4.3, 0.2) &\vline& 0.8488 (1.4, 0.2) && 0.7592 (4.6, 0.0)    \\
1PN   	&\vline& 0.5398 (5.0, 0.1) && 0.5639 (4.3, 0.2) &\vline& 0.5381 (5.0, 0.0) && 0.5614 (4.3, 0.2)    \\
1.5PN 	&\vline& 0.9516 (0.4, 0.2) &&  		        &\vline& 0.9056 (0.4, 0.2) && 		 	   \\
2PN   	&\vline& 0.8751 (0.0, 0.1) && 		        &\vline& 0.9315 (0.0, 0.1) && 			   \\
2.5PN 	&\vline& 0.8517 (0.4, 0.1) && 		        &\vline& 0.8333 (0.0, 0.1) && 			   \\
3PN   	&\vline& 0.9955 (0.0, 0.3) && 		        &\vline& 0.9366 (0.4, 0.1) && 			   \\
3.5PN 	&\vline& 0.9968 (0.0, 0.3) && 		        &\vline& 0.9376 (0.4, 0.1) && 			   \\

\hline
\hline

\label{table:Effectualness2-SC-CM-IL}
\end{tabular}
\end {table}

\begin{table} 
\caption{Faithfulness of the {\it standard adiabatic} $T(E_{[n]},{\cal F}_{n})$ 
and {\it complete adiabatic} $T(E_{[n+2.5]},{\cal F}_{n})$ templates in the 
comparable-mass case. The overlaps are calculated for the initial LIGO noise 
spectrum.}
\begin{tabular}{ccccccccccccccccccc}
\hline
\hline
&\vline
&\multicolumn{5}{c}{$(10M_{\odot},\, 10M_{\odot})$}
&\vline 
&\multicolumn{5}{c}{$(1.4M_{\odot},\, 1.4M_{\odot})$} 
\\
\cline{2-13}
&\vline&\multicolumn{2}{c}{{\it TaylorT1}}&\vline&\multicolumn{2}{c}{{\it TaylorF1}}
&\vline&\multicolumn{2}{c}{{\it TaylorT1}}&\vline&\multicolumn{2}{c}{{\it TaylorF1}}\\
\cline{2-13}
Order ($n$)&\vline& {\it S} & {\it C} &\vline&  {\it S} & {\it C} 
&\vline& {\it S} & {\it C} &\vline&  {\it S} & {\it C}\\
\hline
\hline

0PN   &\vline& 0.5603 & 0.8560 &\vline& 0.5608 & 0.8369 &\vline& 0.3783 & 0.1624 &\vline& 0.4105 & 0.1627 \\
1PN   &\vline& 0.3026 & 0.3491 &\vline& 0.3025 & 0.3502 &\vline& 0.1520 & 0.1615 &\vline& 0.1521 & 0.1614\\
1.5PN &\vline& 0.7949 &        &\vline& 0.7854 &        &\vline& 0.7259 &        &\vline& 0.7094 &       \\
2PN   &\vline& 0.9777 &        &\vline& 0.9601 &        &\vline& 0.5565 &        &\vline& 0.5777 &       \\
2.5PN &\vline& 0.5687 &        &\vline& 0.5686 &        &\vline& 0.5934 &        &\vline& 0.5895 &       \\
3PN   &\vline& 0.9440 &        &\vline& 0.9446 &        &\vline& 0.9888 &        &\vline& 0.9246 &       \\
3.5PN &\vline& 0.9522 &        &\vline& 0.9505 &        &\vline& 0.9916 &        &\vline& 0.9271 &       \\

\hline
\hline
\label{table:Faithfulness-SC-CM-WN}
\end{tabular}
\end {table}

\subsection{ Comparable mass results beyond the adiabatic approximation}
Finally, for the comparable mass case, non-adiabatic waveforms were 
generated in the Lagrangian formalism, using the complete  equations 
Eq.~(\ref{eq:NonAdiabEqnOfMotion}) -- Eq.~(\ref{eq:NonAdiabAccln35pn}). 

Tables~\ref{EffLagrangLIGOCompar} and~\ref{FaithLagrangLIGOCompar} 
show the effectualness and faithfulness of the standard and complete 
non-adiabatic Lagrangian waveforms for the initial LIGO detector. 
The results are more mixed in this case than for the test mass case. 
For the 0PN order, for the NS-NS binary, the standard non-adiabatic 
approach seems to be more effectual and faithful than its complete 
non-adiabatic counterpart. However, in the BH-BH case, the complete 
non-adiabatic seems to be more effectual; but less faithful.  
At the 1PN order, the effectualness is always higher for the complete 
non-adiabatic case; but faithfulness is always lower. 
It is interesting to note that the effectualness trends shown by the 
adiabatic and non-adiabatic approximants are the same at orders 0PN 
and 1PN. However, further work will be necessary to make very strong 
statements in this case.

\begin{table}
\caption{Effectualness of the Lagrangian templates in the comparable mass case for the
initial LIGO noise spectrum. Percentage biases $\sigma_{m}$ and $\sigma_{\eta}$ in 
determining parameters $m$ and $\eta$ are given in brackets. }
\begin{tabular}{ccccccccccccc}
\hline
\hline
& \vline &\multicolumn{3}{c}{$(1.4M_{\odot},1.4M_{\odot})$}
& \vline &\multicolumn{3}{c}{$(10M_{\odot},10M_{\odot})$} \\
\cline{2-9}
Order ($n$)& \vline & {\it S}  && {\it C}  & \vline &  {\it S} && {\it C} \\
\hline
\hline
0PN    & \vline & 0.9282 (27, 1.6) && 0.5848 (32, 3.6) & \vline & 0.8533 (14, 2.5)  && 0.9433 (35, 8.9) \\
1PN    & \vline & 0.5472 (22, 1.3) && 0.6439 (24, 3.1) & \vline & 0.8137 (3.6, 0.2) && 0.9329 (11, 7.4) \\
\hline
\hline
\label{EffLagrangLIGOCompar}
\end{tabular}
\end {table}

\begin{table}
\caption{Faithfulness of the Lagrangian templates in the comparable mass case for the 
initial LIGO noise spectrum.}
\begin{tabular}{ccccccccccccc}
\hline
\hline
& \vline &\multicolumn{2}{c}{$(1.4M_{\odot},1.4M_{\odot})$}
& \vline &\multicolumn{2}{c}{$(10M_{\odot},10M_{\odot})$} \\
\cline{2-7}
Order ($n$)& \vline & {\it S}  & {\it C}  & \vline &  {\it S}  & {\it C} \\
\hline
\hline
0PN     & \vline & 0.0717 & 0.0658  & \vline & 0.6689 & 0.3146 \\
1PN     & \vline & 0.0810 & 0.0771  & \vline & 0.7380 & 0.6568 \\
\hline
\hline
\label{FaithLagrangLIGOCompar}
\end{tabular}
\end {table}

\section{Summary and Conclusion}

The {\it standard adiabatic} approximation to the phasing of gravitational waves 
from inspiralling compact binaries is based on the post-Newtonian expansions 
of the binding energy and gravitational wave flux both truncated at the 
{\it same relative} post-Newtonian order. To go beyond the adiabatic approximation
one  must view the problem as  the dynamics of a  binary under conservative 
relativistic  forces and gravitation radiation damping. In this viewpoint the  
standard  approximation at leading order is equivalent to considering the 0PN 
and 2.5PN terms in the acceleration and neglecting  the intermediate 1PN and 
2PN terms. A complete  treatment of the acceleration at leading order should 
include {\it all} PN terms up to 2.5PN. These define the {\it standard} and 
{\it complete} {\it non-adiabatic} approximants respectively. A new post-Newtonian 
{\it complete adiabatic} approximant  based on  energy and flux functions is 
proposed. At the leading order it  uses the 2PN energy function rather than the 
0PN one in the standard approximation so that heuristically, it does not miss 
any intermediate post-Newtonian terms in the acceleration. We have evaluated 
the performance of the standard adiabatic vis-a-vis complete adiabatic 
approximants, in terms of their {\it effectualness} (i.e. larger overlaps with 
the exact signal) and {\it faithfulness} (i.e. smaller bias in estimation of 
parameters). We restricted our study only to the inspiral part of the signal
neglecting the plunge and quasi-normal mode ringing phases of the binary
\cite{FH98,bd00,dis03,BCV02,DIJS,QNM}. We have studied the problem both for 
the white-noise spectrum and initial LIGO noise spectrum.

The main result of this study is that the conservative corrections to the
dynamics of a binary that are usually neglected in the standard treatment of the
phasing formula are rather important at low PN orders. At the low PN orders, 
they lead to significant improvement in the overlaps between the approximate 
template and the exact waveform. In both  the white-noise and initial LIGO cases
we found that at low ($< $ 3PN) PN orders the effectualness of the approximants 
significantly improves in the complete adiabatic  approximation. However, 
standard adiabatic  approximants of order $\geq$ 3PN  are nearly as good as  
the complete adiabatic approximants for the construction of effectual templates.

In the white-noise case, the {\it faithfulness} of both the approximants 
fluctuates as we go from one PN order to the next and is generally much smaller 
than our target value of 0.965. The fluctuation continues all the way up to 5PN 
order probably reflecting the oscillatory approach of the flux function to the 
exact flux function with increasing PN order. Poor faithfulness also means that 
the parameters extracted using these approximants will be biased. It is again 
interesting to note that complete adiabatic approximants are generally more 
faithful than the standard adiabatic approximants.  For the initial LIGO noise 
case on the other hand,  the faithfulness of the {\it complete adiabatic} 
approximants is vastly better at all orders.  

To the extent possible, we also tried to investigate this problem in the case 
of comparable mass binaries by studying the overlaps of all the approximants 
with a fiducial `exact' waveform using initial LIGO noise spectrum. It is shown 
that, provided we have a T-approximant of the flux function of order $\geq$ 
1.5PN, the {\it standard adiabatic} approximation provides a good lower bound 
to the {\it complete adiabatic} approximation for the construction of both 
{\it effectual} and {\it faithful} templates. This result is in contrast with the 
test mass case where we found that the {\it complete adiabatic} approximation 
brings about significant improvement in {\it effectualness} up to 2.5PN order 
and significant improvement in {\it faithfulness} at all orders.
To achieve the target sensitivity of 0.965 in effectualness, standard adiabatic 
approximants of order 2PN and 3PN are required for the $(10 M_\odot,10 M_\odot)$ 
and $(1.4 M_\odot,1.4 M_\odot)$ binaries, respectively. Whether the complete 
adiabatic approximant achieves this at an earlier PN order is an interesting 
question. It is worth stressing that this result is relevant only for the family of 
inspiral waveforms. In the real physical case of BH-BH binaries the inspiral 
family would not be adequate and must be supplemented by the plunge part of 
waveform as first discussed in \cite{bd00,dis03} and later in \cite{BCV02,DIJS}. 
A discussion of plunge requires a 3PN description of dynamics so that the 2PN 
templates are no longer adequate. This is an example of the second variety of questions 
one can  study in this area referred to in our introduction related to whether a 
template family indeed represents the GWs from a specific astrophysical system.

We have also constructed both standard and complete non-adiabatic approximants 
using the Lagrangian  models in Ref.~\cite{BCV02}. However, we were 
limited in this investigation because of two reasons: Firstly, 
the Lagrangian models are available only up to 3.5PN order, 
and higher order PN accelerations are as yet unavailable which makes it
impossible to calculate the {\it complete non-adiabatic} approximants of order 
$>$ 1PN. Secondly, the only {\it exact} waveform we have has, however been 
constructed only in the {\it adiabatic} approximation. So we are unable to 
make strong statements of general trends and view this effort only as a first step 
towards a more thorough investigation. From the non-adiabatic models studied,
the  conclusion one can draw is that while complete non-adiabatic approximation
improves the effectualness, it results in a decrease in faithfulness.

There is a limitation to our approach which we should point out:
complete adiabatic models can be very well tested in the test mass where
both approximate and exact expressions are available for the various
quantities. However, complete models cannot be worked out to high 
orders in the comparable mass case since they need the energy 
function to be computed to 2.5PN order greater than the flux and
currently the energy function is only known to 3PN accuracy.
Also, due to the lack of an exact waveform, one is constrained to depend
upon some fiducial exact waveform constructed from the approximants themselves.
Though, in the present  paper we have used the new approximants
to construct waveform templates, one can envisage  applications to discuss 
the  dynamics of the binary using numerical integration of the equations 
of motion.

During the course of this study, we also attempted to assess the 
relative importance of  improving the accuracy of the energy function 
and the flux function by systematically studying the approach of the 
adiabatic PN templates constructed with different orders of the energy and the 
flux functions to the exact waveforms. From the study of test-mass templates we 
also conclude that, provided the comparable mass case is qualitatively 
similar to the test mass case, neither the improvement of the accuracy of energy 
function from 3PN to 4PN nor the improvement of the accuracy of flux function from 3.5PN 
to 4PN will result in a significant improvement in effectualness in the comparable 
mass case. 

\acknowledgments
We are grateful to Luc Blanchet and Alessandra Buonanno for discussions and  
comments on the manuscript. This research was supported partly by grants from 
the Leverhulme Trust and the Particle Physics and Astronomy Research Council, UK.  
PA thanks Raman Research Institute and Albert Einstein Institute for 
hospitality and support, and Cardiff University for hospitality during various 
stages of this work. PA also thanks K. G. Arun for useful discussions. 
BRI thanks Cardiff University, IAP Paris and IHES France for hospitality 
during the final stages of the writing of the paper. CAKR thanks Ian Taylor 
and Roger Philip for useful discussions and PPARC for support while BSS thanks 
the Raman Research Institute for supporting his visit in July-August 2004 during 
which some of this research was carried out.

\end {document}